\documentclass[iop,apj,a4paper,numberedappendix,appendixfloats]{emulateapj}
\usepackage{apjfonts, amsmath, amssymb, epsf}
\usepackage{color}

\shortauthors{Homma et al.}
\shorttitle{Chemical evolution model for dSphs}

\begin{document}

\title{A New Chemical Evolution Model for Dwarf Spheroidal Galaxies based on Observed Long Star Formation Histories}

\author{Hidetomo Homma, Takashi Murayama}
\affil{Astronomical Institute, Tohoku University, Aoba-ku, Sendai 980-8578, Japan}

\email{hide@astr.tohoku.ac.jp}

\author{Masakazu A. R. Kobayashi, and Yoshiaki Taniguchi}
\affil{Research Center for Space and Cosmic Evolution, Ehime University, 2-5 Bunkyo-cho, Matsuyama 790-8577, Japan}

\begin{abstract}

 We present a new chemical evolution model for dwarf spheroidal galaxies (dSphs) in the Local Universe.
 Our main aim is to explain both their observed star formation histories and metallicity distribution functions simultaneously.
 Applying our new model for the four local dSphs, that is, Fornax, Sculptor, Leo~II, and Sextans, we find that our new model reproduces the observed chemical properties of the dSphs consistently.
 Our results show that the dSphs have evolved with both a low star formation efficiency and a large gas outflow efficiency compared with the Milky Way, as suggested by  previous works. 
 Comparing the observed $[\alpha / \mathrm{Fe}]$--$\mathrm{[Fe/H]}$ relation of the dSphs with the model predictions, we find that our model favors a longer onset time of Type~Ia supernovae (i.e., 0.5~Gyr) than that suggested in previous studies (i.e., 0.1~Gyr).
 We discuss the origin of this discrepancy in detail.

\end{abstract}


 \section{Introduction}

 In the standard hierarchical galaxy formation models, dwarf galaxies are the most numerous and elemental systems in the universe; larger galaxies are formed from smaller objects like dwarf galaxies through major and/or minor mergers (e.g., \citealt{Kauffmann_93, Cole_94, Robertson_05}).
 Thus, it is important to study these galaxies as the building blocks of normal bright galaxies to understand the formation and evolution of normal galaxies.
 In the Local Group, there are the following two types of dwarf galaxies: dwarf spheroidals (dSphs) and dwarf irregulars (dIrrs) (e.g., \citealt{Mo_10}).
 While dIrrs are gas-rich systems and show current star formation activities, dSphs are gas-poor and consist of mainly old stellar systems.

 Recent observations have unveiled the detailed star formation histories (SFHs) of dwarf galaxies in the local universe \citep{Hernandez_00, Saviane_00, Carrera_02, Rizzi_03, Dolphin_05, Lee_09, Tolstoy_09, Weisz_11, de_Boer_12a, de_Boer_12b}, revealing a significant diversity among the SFHs of dwarf galaxies.
 \citet{Weisz_11} showed color magnitude diagrams (CMDs) of 60 nearby dwarf galaxies within 4~Mpc from our Galaxy using the {\it Hubble Space Telescope} and estimated their SFHs through the CMDs.
 They found that the majority of dwarf galaxies exhibits a long duration of SFH, consisting of both a dominant ancient star formation ($> 10$~Gyr ago) and a lower level of star formation at intermediate times (1--10~Gyr ago).
 They also found that the SFHs between dSphs and dIrrs are generally indistinguishable over most of the cosmic time.
 However, they pointed out that there is a clear difference between them in the most recent $\sim 1$~Gyr; i.e., dSphs show little star formation activity in this epoch, but dIrrs do.

 The SFHs of galaxies also affects their chemical abundance.  
 The stellar chemical abundance in some dSphs has been observationally investigated by using the 8--10~m class large telescopes with high- and/or medium-resolution  optical spectrographs \citep{Shetrone_01, Shetrone_03, Tolstoy_03, Sadakane_04, Geisler_05, Bosler_07, Aoki_09, Cohen_09, Cohen_10, Kirby_10, Letarte_10, Tafelmeyer_10, Aden_11, Starkenburg_13}.
 The derived chemical abundance indicates that dSphs have fewer metal-rich stars than the Milky Way (MW) disk; the average metallicity of stars in dSphs are $\sim 10$--100 times lower than that of the MW disk stars.
 Moreover, their chemical abundance show different trends from the MW halo, which suggests that dSphs have experienced different chemical evolution histories from that of the MW.

 The chemical evolution histories of the dSphs have also been investigated theoretically based on the numerical models \citep{Ikuta_02, Carigi_02, Lanfranchi_03, Lanfranchi_04, Fenner_06, Cescutti_08, Calura_09, Kirby_11a, Tsujimoto_11, Romano_13}.
 These studies have revealed that both gas outflow and infall are important to reproduce the observed chemical abundance and/or the metallicity distribution functions (MDFs) of dwarf galaxies (\citealt{Lanfranchi_04, Kirby_11a}, hereafter K11).
 For dSph, a large gas outflow efficiency and lower star formation efficiency (SFE) than that of the MW have been predicted for their small metallicities; the outflow rate per star formation rate (SFR) and the SFE of the MW are $\sim 0.5$ and $\sim 0.24~\mathrm{Gyr^{-1}}$, respectively \citep{Joung_12, Kennicutt_12}.

 Recently, \citet{Kirby_10} measured chemical abundance of 2961 stars in eight dSphs, which provided a large and homogeneously analyzed sample of stellar spectra in dSphs.
 In order to extract SFHs of the dSphs from their observed data, K11 constructed a detailed chemical evolution model including chemical yields from Types~Ia and II supernovae (SNe~Ia and SNe~II, respectively) and from asymptotic giant branch stars (AGBs) as well as gas infall and outflow induced as SN wind.
 The model simultaneously reproduced the observed trends of $\mathrm{[Mg/Fe]}$, $\mathrm{[Si/Fe]}$, and $\mathrm{[Ca/Fe]}$ ratios with $\mathrm{[Fe/H]}$ as well as MDFs in the dSphs.
 However, it should be emphasized that the duration timescales of their derived SFHs were only $\sim 1$~Gyr (see Table~3 in K11), which are much shorter than those observationally estimated from the CMDs in dSphs \citep{Dolphin_02, Lee_09, de_Boer_12a, de_Boer_12b}.
 Moreover, although they stated that the derived star formation timescales were extremely sensitive to the minimum delay time of SN~Ia, $t_\mathrm{delay,min}$, they showed the resultant SFHs adopting two different values of $t_\mathrm{delay,min}$ only for a dSph among their eight dSphs in total (see Figure~10 in K11).

 In this paper, we construct a new analytic model of the chemical evolution of galaxies to reproduce both the observed SFHs and MDFs of dSphs, simultaneously.
 Our model explicitly utilizes the SFH derived from observed CMD in dSph as the most fundamental input quantity.
 Although this idea is similar to that proposed by \citet{Ikuta_02} and \citet{Carigi_02}, they compared their results only with the observed chemical abundance ratios of stars in dSphs.
 Since MDF is more sensitive to SFH than abundance ratio, we compare our model results with the observed MDFs in dSphs.
 While the abundance ratios are determined by the yields of chemically dominant stars at each epoch, MDF depends on both SFH and age-metallicity relation (AMR).
 Therefore, if we find a consistent solution for both CMD and chemical property of a certain dSph, we obtain its MDF and AMR simultaneously from our chemical evolution model based on its observationally estimated SFH.
 We also compare our model results for the $\mathrm{[Mg/Fe]}$--$\mathrm{[Fe/H]}$ relation to investigate the minimum delay time of SN~Ia; since the SN Ia explodes with a delay time and ejects large amount
 of Fe relative to the SN~II, the $\mathrm{[Mg/Fe]}$--$\mathrm{[Fe/H]}$ relation suddenly changes when the SN~Ia starts to explode.

 The rest of this paper is organized as follows.
 The details of our chemical evolution model are described in Section~\ref{sec:model}.
 In Section~\ref{sec:results}, we present the results of our model in comparison with observations.
 After some discussions including the comparison of our model results with K11's one in Section~\ref{sec:discussion}, the summary of this paper is given in Section~\ref{sec:summary}.

 \section{The Chemical Evolution Model Calculations}\label{sec:model}

  \subsection{Basic Assumption and Formulation}\label{sec:assumption}

  Here we describe our new chemical evolution model to reproduce simultaneously both the observed SFHs and MDFs of the following four local dSphs, i.e., Fornax, Sculptor, Leo~II, and Sextans.
  We adopt the following assumptions in our model, which are the same as those in K11 except the items 2, 6, and 8.
  \begin{enumerate}
   \item The interstellar medium (ISM) of a dSph is treated as one zone; that is, we do not consider any spatial variations of gas density and metallicity in the ISM at any time.
   \item The observed SFH derived from the analysis of CMD is used in our model. Therefore, the SFR at time $t$, $\Psi(t)$, is fixed to the observed one.
   Namely, our model is  calculated along with the SFH.\label{assump:SFR}
   \item We assume that there is a positive correlation between the gas mass and SFR like the Schmidt-Kennicutt law \citep{Schmidt_59, Kennicutt_98}.
   Since the SFR is given as input, the gas mass at a certain time is completely determined according to the SFR at the time.\label{assump:Mgas}
   \item The interstellar gas can be expelled from dSph (i.e., gas outflow) while supplied from its outside (i.e., gas infall); that is, the dSph in our model is not a closed-box.\label{assump:open}
   \item Gas outflow is assumed to be induced via a superwind (a collective effect of a large number of SNe), so that its rate is proportional to the total number of SNe~Ia and II at each time.
   Metallicity of the outflow gas is assumed to be the same as that of the system at that time.\label{assump:outflow}
   \item The number of SN~Ia is determined by the past SFH and the delay time distribution (DTD) function.
   The DTD function used in our model is the observationally derived one of \citet{Maoz_10}.
   Minimum delay time of SN~Ia, $t_\mathrm{delay,min}$, is adopted to be one of the model parameters.\label{assump:SNIa}
   \item The number of SN~II is determined by the past SFH and the stellar lifetime given by \citet{Timmes_95}.
   We assume that stars with $m \ge 10~M_\odot$ explode as SNe~II.\label{assump:SNII}
   \item Gas infall rate is adjusted to reproduce the observed SFH.
   Since we assume that the SFR is fixed (item~\ref{assump:SFR}) and the gas mass of the ISM correlates with the SFR (item~\ref{assump:Mgas}), the gas mass at time $t$, $M_\mathrm{gas}(t)$, is determined by the observed SFH.
   Moreover, the time derivative of the gas mass, $\dot{M}_\mathrm{gas}(t)$, depends on the SFR, the ejected gas mass from evolved stars (i.e., AGBs, and SNe~Ia and II), the gas outflow via a superwind (item~\ref{assump:outflow}), and the gas infall.
   Therefore, in order to reproduce the gas mass, we set the gas infall rate as an adjustable amount (see Section~\ref{sec:infall}).
   Metallicity of the infall gas is assumed to be primordial.\label{assump:infall}
   \item The yield of SNe~Ia and II is adopted from \citet{Iwamoto_99} and \citet{Nomoto_06}, respectively.
   For the AGB yields, we adopt those of \citet{Karakas_10}.
   We assume that all stars with $m <$ $10~M_{\odot}$ evolve into AGBs after their lifetimes.
   All heavy elements are assumed to stay in gas-phase during the entire time of calculation.
   \item The initial mass function (IMF) used in our model, $\phi (m)$, is that derived by \citet{Kroupa_93} with a mass range of $m = 0.08$--$100~M_{\odot}$.
  \end{enumerate}

  We calculate the chemical evolution for 119 elements and isotopes listed in \citet{Iwamoto_99}, \citet{Nomoto_06}, and \citet{Karakas_10} under the assumptions listed above.
  The equation of the mass evolution and the chemical evolution for an element $i$ at time $t$ are represented by
  \begin{eqnarray}
   \dot{M}_\mathrm{gas}(t) &=& -\Psi(t) + E(t) + \dot{M}_\mathrm{in}(t) - \dot{M}_\mathrm{out}(t) \label{eq:mass_evolution}\\
   \mathrm{and}\quad \dot{M}_{\mathrm{gas},i}(t) &=& -\Psi(t)X_i(t) + E_i(t) + \dot{M}_{\mathrm{in}}(t)X_{\mathrm{in},i}(t) \nonumber\\
    & &\quad - \dot{M}_\mathrm{out}(t)X_i(t),\label{eq:CEM}
  \end{eqnarray}
  where $M_{\mathrm{gas},i}(t)$ is the gas-phase mass of an element $i$ in the ISM at time $t$, $X_i(t)$ is the gas-phase mass fraction for the element to the total mass of interstellar gas ($X_i (t) \equiv M_{\mathrm{gas},i} (t) / M_\mathrm{gas} (t)$ and $\sum_i X_i (t) = 1$), $E_i(t)$ is the rate of gas mass for the element ejected from evolved stars to the ISM and $E(t)$ is its total ($E(t) = \sum_i E_i(t)$), and $\dot{M}_{\mathrm{in}}(t)$ and $\dot{M}_{\mathrm{out}}(t)$ are its infall and outflow rates, respectively.
  We assume that both the initial abundance of the ISM and the infall gas are primordial (i.e., $X_\mathrm{H} (t = 0) = X_\mathrm{in,H} = 0.75$ and $X_\mathrm{He} (t = 0) = X_\mathrm{in,He}$ $= 0.25$).

  The ejected gas mass $E_i(t)$ is given by the sum of three independent terms as follows:
  \begin{equation}
   E_i(t) = \dot{\xi}_{i,\mathrm{Ia}}(t) + \dot{\xi}_{i,\mathrm{II}}(t) + \dot{\xi}_{i,\mathrm{AGB}}(t),\label{eq-Ei}
  \end{equation}
  where $\dot{\xi}_{i,\mathrm{Ia}} (t)$, $\dot{\xi}_{i,\mathrm{II}} (t)$, and $\dot{\xi}_{i,\mathrm{AGB}} (t)$ represent the rates of gas masses ejected from SN~Ia \citep{Iwamoto_99}, SN~II \citep{Nomoto_06}, and AGB \citep{Karakas_10}, respectively.
  The term $\dot{M}_{\mathrm{out}}(t)$ can be written by using the number of SNe~II and Ia, $\dot{N}_\mathrm{II}(t)$ and $\dot{N}_\mathrm{Ia}(t)$, respectively, as
  \begin{eqnarray}
   \dot{M}_\mathrm{out}(t) &=& A_\mathrm{out} \left[\dot{N}_\mathrm{Ia}(t) + \dot{N}_\mathrm{II}(t) \right], \label{eq-Mout}\\
   \mathrm{where}\quad \dot{N}_\mathrm{Ia}(t) &=& \int^{t}_{0} {\it DTD}(t-t') \Psi(t')\ dt',\\
   \quad \mathrm{and}\quad \dot{N}_\mathrm{II}(t) &=& \int^{100~M_\odot}_{10~M_\odot} \Psi(t - \tau (m)) \phi(m)\ dm \label{eq-dotNII}.
  \end{eqnarray}
  Here the outflow coefficient $A_\mathrm{out}\ (M_\odot~\mathrm{SN}^{-1})$ controls the outflow gas mass per unit SN.
  For the DTD function, we adopt the following one, which are observationally estimated by \citet{Maoz_10}:
  \begin{equation}
   {\it DTD}(t_\mathrm{delay}) =
    \begin{cases}
     1 \times 10^{-3}\ \mathrm{SN\ Gyr}^{-1}\ M_{\odot}^{-1}\\
     \qquad \times \left(t_\mathrm{delay} / \mathrm{Gyr}\right)^{-1.1} & \mathrm{for}\ t_\mathrm{delay} \ge t_\mathrm{delay,min}\\
     0 & \mathrm{otherwise}
    \end{cases}
    ,
  \end{equation}
  where $t_\mathrm{delay}$ indicates an elapsed time since the progenitor star of SN Ia is formed.
  The variable $\tau(m)$ in Equation~(\ref{eq-dotNII}) is the stellar lifetime of the star with mass of $m$.

  The interval of each time step in our calculation is $\Delta t = 25$~Myr\footnote{This time interval is longer than that of K11, $\Delta t_\mathrm{K11} = 1$~Myr.
  We confirm that this difference of $\Delta t$ does not affect the results significantly (see Appendix).}.  
  The calculation is terminated when the SFR reaches zero or a certain condition, which is described in detail in Section~\ref{sec:likelihood}, is satisfied.

  The assumptions adopted in our model listed above are the same as K11 except the assumptions for the SFH (item~\ref{assump:SFR}), minimum delay time of SN Ia (item~\ref{assump:SNIa}), and gas infall rate (item~\ref{assump:infall}).
  It should be stated that, while the stellar lifetime adopted in our model \citep{Timmes_95} is also different from that in K11 (i.e., \citealt{Padovani_93, Kodama_97}), the difference is found to not affect the resultant chemical evolution.
  In the following subsections, we describe these major differences in the model assumptions between K11 and ours.

   \subsubsection{Star Formation History and Gas Mass}\label{subsubsec:SFH}

\begin{figure}
    \begin{center}
     \plotone{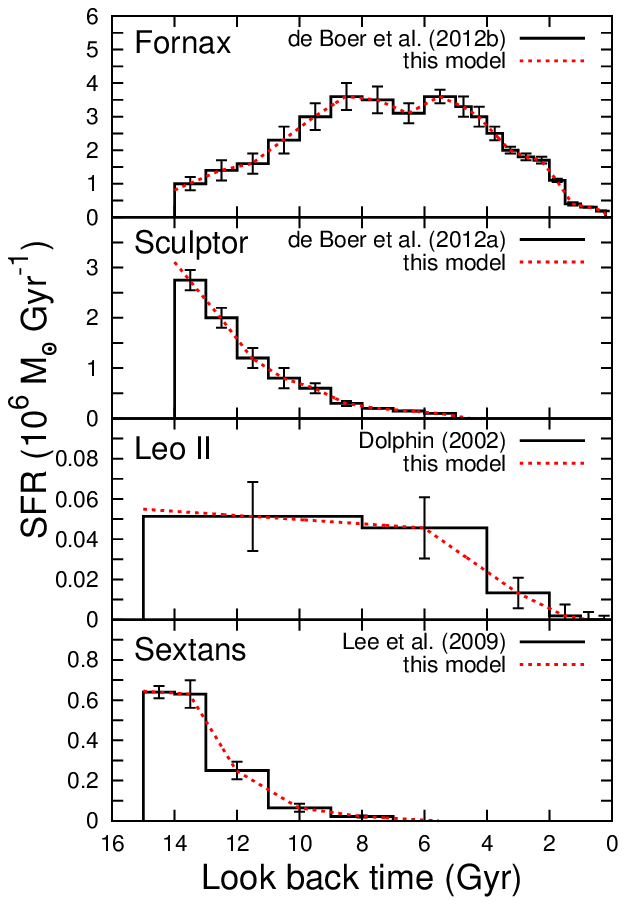}
     \caption{The SFHs of our sample dSphs.
	 The histograms with error bars are observational estimates for the SFHs derived from the CMDs (Fornax: \citealt{de_Boer_12b}, Sculptor: \citealt{de_Boer_12a}, Leo~II: \citealt{Dolphin_02}, Sextans: \citealt{Lee_09}).
	 The error bars indicate the uncertainties caused by the so-called CMD model by using which the SFH can be estimated from the CMD.
	 The red dotted lines are the SFHs used in our model calculations obtained through a linear interpolation of the observed SFHs.\label{fig:SFH}}
    \end{center}
   \end{figure}

   In our model, we use the SFHs estimated from the photometric observations for our sample dSphs (i.e., Fornax, Sculptor, Leo~II, and Sextans) as the most fundamental inputs.  
   On the other hand, in K11, SFH is not an input quantity but one of the output quantities, which is determined to reproduce the observed data of the MDF and abundance ratios for each one of their samples of the local dSphs including our sample dSphs.
   In Figure~\ref{fig:SFH}, we show the SFHs of our sample dSphs derived in literature (Fornax: \citealt{de_Boer_12b}, Sculptor: \citealt{de_Boer_12a}, Leo~II: \citealt{Dolphin_02}, Sextans: \citealt{Lee_09}).
   It should be emphasized that the observed SFHs have much longer timescales of about 8 to 13~Gyr than the timescales of the SFHs derived in K11 of $\sim 0.8$--1.3~Gyr.
   
   It should be noted that the SFHs of both Leo~II and Sextans start from 15~Gyrs ago, which is longer than the age of the universe derived from the current cosmology \citep{Komatsu_09}.
   As for this discrepancy, \citet{Lee_09} mentioned that the current level of discrepancy is within the range of various plausible adjustments for the stellar models.
   However, for the safety, we have compared the best-fit parameters, whose details are  described in Section \ref{sec:likelihood}, of Sextans for the SFHs starting from 15~Gyrs ago and 14~Gyrs ago and we have confirmed that there is no difference.
   Namely, the discrepancy is not critical to the chemical evolution of the dSphs.
   Therefore we have decided to let it stand.

   For each dSph in our sample, we derive its SFR at each time step via linear interpolation of the discrete data points of its observed SFH.
   Note that these observational estimates for SFH used here contain uncertainties, as shown in Figure~\ref{fig:SFH}, originated from the model which decodes the SFH of a dSph from its CMD, so-called ``CMD model''.
   However, we do not incorporate such uncertainties in the process to obtain $\Psi (t)$ via linear interpolation; therefore, SFH used in our calculation does not include uncertainties of the CMD model.

   We then calculate the gas-phase mass in the ISM of the dSph at time $t$, $M_\mathrm{gas}(t)$, as
   \begin{equation}
    \frac{M_{\mathrm{gas}}(t)}{10^6~M_\odot} = \left[\frac{1}{A_\ast}\left(\frac{\Psi(t)}{10^6~M_\odot~\mathrm{Gyr}^{-1}}\right)\right]^{1/\alpha}, \label{eq:mass}
   \end{equation}
   where $A_\ast$ and $\alpha$ are the same parameters in the K11 model.
   While K11 defined the unit of $A_\ast$ as $10^6~M_\odot~\mathrm{Gyr}^{-1}$, we set $A_\ast$ as the non-dimension parameter.
   This formulation explains the relation between total mass of interstellar gas $M_\mathrm{gas}(t)$ and $\Psi(t)$ (i.e., $\Psi \propto M_\mathrm{gas}^\alpha$, the generalization of a Kennicutt-Schmidt law).
   While K11 treated $\alpha$ as one of their model parameters and found that its best-fit value varies among their sample of dSphs (see Table~2 in K11), we fix it to unity in our model for simplicity.
   When $\alpha=1$, the parameter $A_\ast$ takes the same value of the SFE (i.e., $=\Psi(t)/M_\mathrm{gas}(t)$) in unit of $\mathrm{Gyr}^{-1}$.
   We investigate the effect of using a fixed $\alpha$ on other model parameters through a comparison with the K11 model results in Section~\ref{sec:K11}.

   \subsubsection{Type Ia Supernova}

   We assume that SNe~Ia explode according to the past SFH and the adopted DTD function with the parameter which represents the minimum delay time for SN~Ia, $t_\mathrm{delay, min}$.
   These assumptions are the same as that of K11 except for the value of $t_\mathrm{delay,min}$.
   K11 adopted a single value of $t_\mathrm{delay, min} = 0.1$~Gyr, while they concluded that the derived star formation timescale is extremely sensitive to $t_\mathrm{delay, min}$; larger $t_\mathrm{delay, min}$ results in longer star formation timescale.
   Since the star formation timescale in our model is much longer than that derived in K11, we adopt the following two values for the minimum delay time for SN~Ia: $t_\mathrm{delay, min} = 0.1$~Gyr and 0.5~Gyr.

   \subsubsection{Gas Infall}\label{sec:infall}

   While the gas infall rate at a certain time $t$, $\dot{M}_\mathrm{in}(t)$, is represented by $\dot{M}_\mathrm{in} (t) \propto t e^{-t / \tau_\mathrm{in}}$ in K11, where $\tau_\mathrm{in}$ is one of their model parameters, it is determined in our model so that Equation~(\ref{eq:mass_evolution}) is satisfied at any time in calculation.
   This difference comes from the different assumption for the SFH described in Section~\ref{subsubsec:SFH}.
   In our model, once the SFH of a certain dSph is given, $M_{\mathrm{gas}} (t)$ at any $t$ is determined via Equation~(\ref{eq:mass}) with an adopted parameter of $A_\ast$ and hence $\dot{M}_{\mathrm{gas}} (t)$ is also fixed.
   Moreover, $\dot{\xi}_{i,\mathrm{II}} (t)$ and $\dot{\xi}_{i,\mathrm{AGB}} (t)$ in Equation~(\ref{eq-Ei}) and $\dot{N}_\mathrm{II} (t)$ in Equation~(\ref{eq-Mout}) are completely determined because the stellar lifetime and both yields of SN~II and AGB used in our model are all fixed.
   The terms $\dot{\xi}_{i,\mathrm{Ia}} (t)$ and $\dot{N}_\mathrm{Ia} (t)$ in these equations also depend on SFH as well as one of the parameters in our model, $t_\mathrm{delay,min}$.
   Therefore, all variables in Equation~(\ref{eq:mass_evolution}) except for $\dot{M}_\mathrm{in}(t)$ are determined for a given SFH and set of parameters in our model, that is, $A_\ast$, $A_\mathrm{out}$, and $t_\mathrm{delay, min}$.
   In more qualitative sense, our prescription for gas infall rate means that, while SN blows out a part of gas from the ISM of dSph, star formation can take place according to the observed SFH so that the primordial gas flows into the system at the same time.

   Nevertheless, the gas infall rate determined in this way would be negative when the time derivative of gas mass, $\dot{M}_{\mathrm{gas}} (t)$, is negative and the outflow rate, $\dot{M}_\mathrm{out} (t)$, is low.
   Such negative infall rate acts as outflow effectively.
   However, this implies that the system would blow out not only enriched gas by SNe but also extra primordial gas simultaneously.
   This is against to one of the assumptions in our model that the system is chemically homogeneous.
   To avoid such physically unlike situation, we artificially add an extra outflow rate $\dot{M}_\mathrm{out}^\mathrm{ex}$ to Equation~(\ref{eq:mass_evolution}) so that the resulting infall rate becomes zero during such situation.
   Therefore, the gas infall rate is determined via the following equation rather than Equation~(\ref{eq:mass_evolution}):
   \begin{equation}
    \dot{M}_\mathrm{in}(t) = \dot{M}_\mathrm{gas}(t) + \Psi(t) - E(t) + \dot{M}_\mathrm{out}(t) + \dot{M}^\mathrm{ex}_\mathrm{out}(t) .\label{eq:sec2-2}
   \end{equation}
   Since the metallicity of outflow gas is assumed to be the same as that of the system at that time in our model, introducing such extra outflow gas solve the physically unlike situation described above.
   Although the introduction of such extra outflow rate is completely artificial, it should be stated that external processes to remove gas from dSph such as tidal stripping can act as this extra outflow rate.
   This is because the outflow rate in our model is simply assumed to be linearly proportional to the SN rates as represented by Equation~(\ref{eq-Mout}).

   As a consequence of introducing the extra outflow rate, the net outflow rate ($= \dot{M}_\mathrm{out} (t) + \dot{M}_\mathrm{out}^\mathrm{ex} (t)$) is not proportional to the SN rate transiently when $\dot{M}_\mathrm{out}^\mathrm{ex}$ is not equal to zero.
   In order to evaluate the importance of the extra outflow rate quantitatively, we define $F_\mathrm{ex}$ as the total mass ratio of the extra outflow to the net outflow in the entire SFH of a dSph, which can be represented by
   \begin{equation}
    F_\mathrm{ex} = \int^{t_\mathrm{fin}}_0 \dot{M}^\mathrm{ex}_\mathrm{out} (t) dt \bigg/ \int^{t_\mathrm{fin}}_0 \left(\dot{M}_\mathrm{out} (t) + \dot{M}_\mathrm{out}^\mathrm{ex} (t) \right) dt,
   \end{equation}
   where $t_\mathrm{fin}$ is the time at which our calculation ends (see Section~\ref{sec:likelihood} for its detailed definition).
   If $F_\mathrm{ex} = 0$, there is no extra outflow gas required for the dSph during model calculation.
   If $0 < F_\mathrm{ex} \le 1$, the model needs extra gas outflow during model calculation.

   Although this modification is artificial, the derived best-fit model parameters are consistent within their $1\sigma$ uncertainties whether or not this modification is adopted for our sample.
   The detailed definition of the best-fit model and $1\sigma$ uncertainties are described in next section and Section~\ref{sec:best-fit}, respectively.

  \subsection{Determination of the Best-fit Model} \label{sec:likelihood}

  In our model, there are three free parameters to fit the observed MDF as follows: SFE ($A_\ast$), outflow coefficient ($A_\mathrm{out}$), and minimum delay time for SN~Ia ($t_\mathrm{delay,min}$).
  The best-fit parameters are determined through the following two steps.
  First, for each dSph in our sample, we calculate the evolution history of iron abundance, $\zeta_\mathrm{Fe}(t) \equiv \mathrm{[Fe/H]}(t)$, using the model provided with its SFH.
  When we calculate the abundance, we adopt the solar composition of \citet{Sneden_92} for Fe; $12 + \log(n_\mathrm{Fe} / n_\mathrm{H})_\odot = 7.52$.
  The parameter ranges explored are $A_\ast = 10^{-3}$--$10$ and $A_\mathrm{out} = 10^2$--$10^5~M_{\odot}~\mathrm{SN}^{-1}$, equally spaced in logarithmic scale with a step of 0.01~dex.
  For $t_\mathrm{delay,min}$, we examine the following two cases: $t_\mathrm{delay,min} = 0.1$~Gyr and 0.5~Gyr.
  Therefore, the total number of the parameter sets is 120,000 for each of $t_\mathrm{delay,min}$.
  Then, the best-fit model for each one of $t_\mathrm{delay,min}$ is individually determined by using a maximum likelihood estimation, which maximizes the following likelihood $L$:
  \begin{eqnarray}
   L &=& \prod_j \int^{t_\mathrm{fin}}_0 \frac{1}{\sqrt{2\pi}\ \Delta \zeta_{\mathrm{Fe},j}}\ \exp{\left[-\frac{\left(\zeta_{\mathrm{Fe},j} - \zeta_\mathrm{Fe}(t)\right)^2}{2(\Delta \zeta_{\mathrm{Fe},j})^2}\right]} \nonumber \\
   && \qquad \qquad \times \ \frac{\dot{M}_*(t)}{M_*}\ dt,\label{eq:likelihood}
  \end{eqnarray}
  where $\zeta_{\mathrm{Fe},j} = \mathrm{[Fe/H]}_j$ and $\Delta \zeta_{\mathrm{Fe}, j} = \Delta \mathrm{[Fe/H]}_j$ are the observed iron abundance and its uncertainty of the $j$th star in the dSph, respectively.
  The variable $\dot{M}_\ast (t) / M_\ast$ is evaluated in our model according to the SFH of the dSph, where $M_*$ is the stellar mass of the dSph at present and $\dot{M}_*(t)$ is the mass of the stars that are newly formed in the dSph at time $t$ and survive until the present.
  We note that, since both $M_\ast$ and $\dot{M}_\ast(t)$ are completely determined from the input SFH, the factor $\dot{M}_\ast (t) / M_\ast$ does not depend on our model parameters and simply acts as a weight in the likelihood calculation.
  Therefore, the best-fit parameters can be different among our sample dSphs.

  For the observed iron abundance of our sample dSphs, we use the data provided in \citet{Kirby_10}, in which the iron abundance $[\mathrm{Fe/H}]$ of nearly 3,000 stars in eight dSphs including our sample dSphs is spectroscopically measured by using Fe~I absorption lines.
  It should be noted that these sample stars are different from those used to derive the SFHs.
  In general, the area in a dSph spectroscopically observed is different from that photometrically observed as shown in Table~\ref{tab:size}; the area observed spectroscopically is typically smaller than that observed photometrically and concentrates on the center of each dSph.
  However, since we assume that the dSph in our model is chemically homogeneous at any time, we simply assume that the stars sampled in the photometric and spectroscopic observations both well represent the entire population in the dSph.
  We will discuss the influences of such difference in the observed area on the resultant chemical evolution histories in Section~\ref{sec:4-2}.
  \begin{deluxetable}{lcccc}
   \tablewidth{\linewidth}
   \tablecolumns{5}
   \tablecaption{Half light radii and observed area of the dSphs}
   \tablehead{
   & & \multicolumn{2}{c}{Observed area} & \\
   \cline{3-4}
   \colhead{dSph}     & \colhead{Half light radius} & \colhead{Spectroscopy} & \colhead{Photometry} & \colhead{Reference}
   }
   \startdata
   Fornax   & $\sim 16^\prime$   & $\sim 20^\prime \times 20^\prime$ & $\sim 1\fdg 5 \times 1\fdg 0$ & 1, 2, 3 \\
   Sculptor & $\sim 11^\prime$   & $\sim 20^\prime \times 20^\prime$ & $\sim 1\fdg 7 \times 1\fdg 7$ & 1, 2, 4 \\
   Leo~II   & $\sim 2\farcm 6$  & $\sim 20^\prime \times 15^\prime$ & $\sim 1\farcm 3 \times 1\farcm 3$ & 1, 2, 5 \\
   Sextans  & $\sim 28^\prime$   & $\sim 35^\prime \times 20^\prime$ & $\sim 42^\prime \times 28^\prime$ & 1, 2, 6
   \enddata
   \label{tab:size}
   \tablerefs{(1) \citet{McConnachie_12}; (2) \citet{Kirby_10}; (3) \citet{de_Boer_12b}; (4) \citet{de_Boer_12a}; (5) \citet{Mighell_96}; (6) \citet{Lee_09}}
  \end{deluxetable}

  In our calculation for chemical evolution, the metallicity of the system can change violently with time.
  Specifically, such violent change of metallicity happens just before the end of the SFH, where the mass of interstellar gas becomes very low and comparable with that of outflowing gas.
  In this condition, small amount of yield is sufficient to change the metallicity of the interstellar gas, which is supplied via infall of gas with primordial composition.
  Therefore, we set the following two criteria to determine the condition for the violent change of metallicity;
  \begin{eqnarray}
   (1) && \lvert \zeta_\mathrm{Fe}(t - \Delta t) - \zeta_\mathrm{Fe}(t) \rvert > 0.3 \ \mathrm{and} \nonumber \\
   && \quad \left[\zeta_\mathrm{Fe}(t - \Delta t) - \zeta_\mathrm{Fe}(t)\right] \nonumber \\
   && \qquad \times \ \left[\zeta_\mathrm{Fe}(t) - \zeta_\mathrm{Fe}(t + \Delta t)\right] < 0,\\
   \mathrm{or} && \nonumber \\
   (2) &&\vert \zeta_\mathrm{Fe}(t - \Delta t) - \zeta_\mathrm{Fe}(t) \vert > 0.3 \ \mathrm{and} \ t > 5~\mathrm{Gyr},
  \end{eqnarray}
  where $\Delta t$ is the interval of individual time step ($= 25$~Myr).
  If either of these criteria are satisfied, we terminate the calculation and set $t_\mathrm{fin}$ at the time, which is used to calculate the likelihood via Equation~(\ref{eq:likelihood}).

 \section{Results}\label{sec:results}

  \subsection{The Best-fit Models}\label{sec:best-fit}

  \begin{figure*}
   \begin{center}
    \plotone{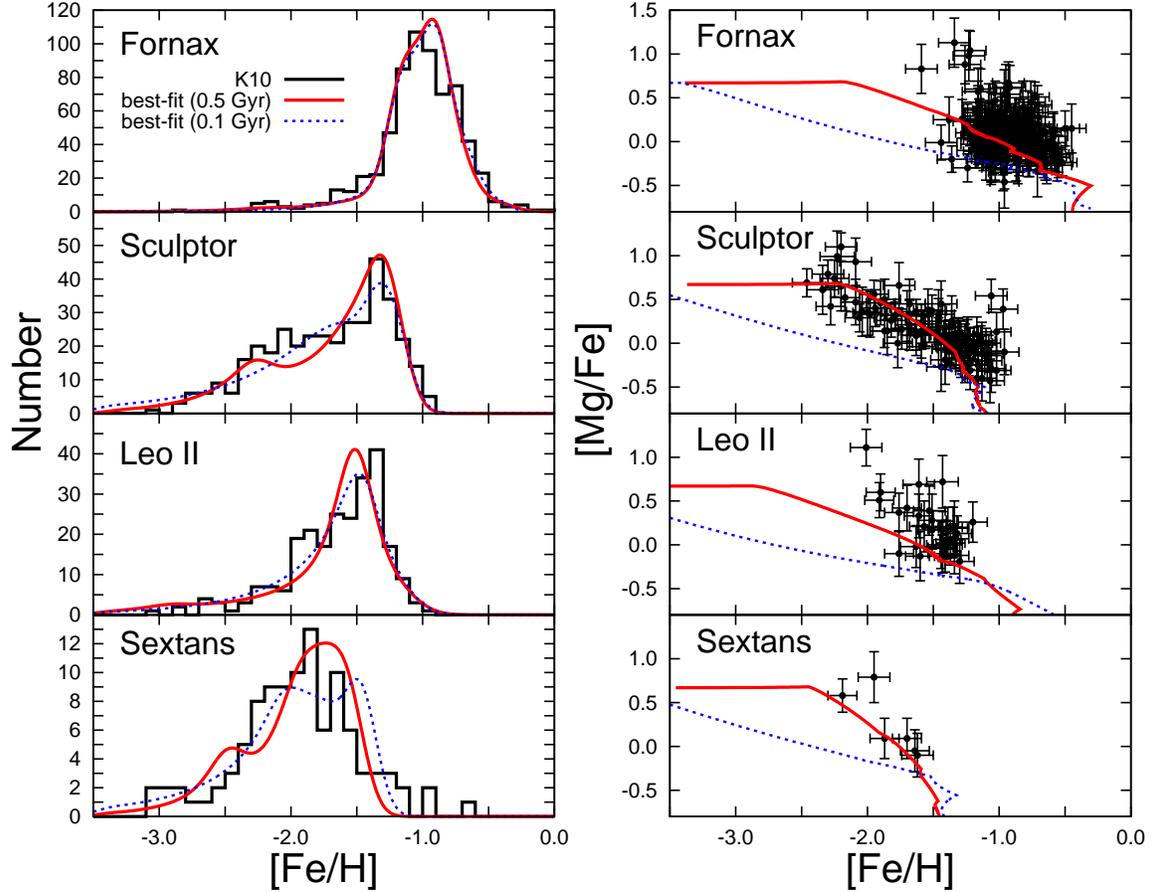}
    \caption{MDFs and $\mathrm{[Mg/Fe]}$--$\mathrm{[Fe/H]}$ diagrams.
	The histograms (left) and data points with error bars (right) show the observed MDFs and $\mathrm{[Mg/Fe]}$ ratios, respectively, taken from \citet{Kirby_10}.
	Only observed data with uncertainties less than or equal to 0.3~dex are shown.
	The blue dotted and red solid curves are the results of our best-fit models with $t_\mathrm{delay,min} = 0.1$~Gyr and 0.5~Gyr, respectively.\label{fig:MDF}}
   \end{center}
  \end{figure*}
  We first compare the MDFs and chemical abundance ratios of the best-fit models for our sample dSphs with those of the observed data in Figure~\ref{fig:MDF}.
  For the observed data in both left and right panels of Figure~\ref{fig:MDF}, we plot only the stars with the uncertainties both in metallicity and abundance ratio less than or equal to 0.3~dex while we do not adopt any cutoff for the uncertainties during the likelihood calculation.
We also show the abundance ratios of some of the other elements calculated in our model, that is, $\mathrm{[Si/Fe]}$, $\mathrm{[Ca/Fe]}$, and $\mathrm{[Ti/Fe]}$ in Appendix.

  It is found that the resultant MDFs of the best-fit models reasonably reproduce the observed data.
  Specifically, the peak metallicities at which the observed MDFs have the largest frequencies of stars are well fitted by the best-fit models for all of our sample dSphs.
  Moreover, the tails in the observed MDFs elongated into metal-poor sides from the peak metallicities, which are different among the dSphs, are also reproduced by the best-fit models.
  On the other hand, although the observed MDF for Sextans shows a broader metal-rich tail with $\Delta [\mathrm{Fe/H}] \sim 1$~dex than those for other three dSphs of $\Delta [\mathrm{Fe/H}] \sim 0.5$~dex, the model MDFs have similar metal-rich tails with $\Delta [\mathrm{Fe/H}] \sim 0.5$~dex in all dSphs despite the difference in their SFHs.
  It is also found that there are bumps in the model MDFs with $t_\mathrm{delay,min} = 0.5$~Gyr at $\mathrm{[Fe/H]} \sim -2.3$ in Sculptor and $\mathrm{[Fe/H]} \sim -2.5$ in Sextans while such bumps are not evident in the observed MDFs.

  The peak metallicities in the MDFs are found to correlate with $A_\mathrm{out}$; the model MDF with smaller $A_\mathrm{out}$ has a peak at higher metallicity.
  It is also found that the metal-poor tail is related to SFH; the dSph forming stars more intensively in the past tends to have more elongated tail.
  On the other hand, the significance of the bump seen in the metal-poor tail is found to correlate with the model parameters $A_\ast$ and $t_\mathrm{delay,min}$ as well as the SFH.
  We discuss the effects of our model parameters of $A_\mathrm{out}$, $A_\ast$, and $t_\mathrm{delay,min}$ on the MDF are discussed in detail in Section~\ref{sec:4-1}.

  \subsection{The Best-fit Model Parameters and Their Uncertainties}\label{subsec:Param+Errors}

  \begin{deluxetable*}{lccccccccc}
   \tablecolumns{8}
   \tablecaption{The best-fit model parameters with their $1\sigma$ uncertainties, the maximum likelihoods, and the fractions of extra outflow gas mass}
   \tablehead{
    \parbox[c][17pt][c]{0cm}{}  &
   \multicolumn{4}{c}{$t_\mathrm{delay,min} = 0.1$~Gyr} & & \multicolumn{4}{c}{$t_\mathrm{delay,min} = 0.5$~Gyr} \\
   \cline{2-5} \cline{7-10}
   \parbox[c][17pt][c]{0cm}{} dSph & $\log{A_\ast}$ & $\log{[A_\mathrm{out}/(M_{\odot}~\mathrm{SN}^{-1})]}$ & $-\log{L_\mathrm{max}}$ & $F_\mathrm{ex}$ (\%) & & $\log{A_\ast}$ & $\log{[A_\mathrm{out}/(M_{\odot}~\mathrm{SN}^{-1})]}$ & $-\log{L_\mathrm{max}}$ & $F_\mathrm{ex}$ (\%)
   }
   \startdata
   \parbox[c][17pt][c]{0cm}{} Fornax & $-1.50^{+0.05}_{-0.03}$ & $3.26^{+0.02}_{-0.02}$ & 175 & 4.4 & & $-1.01^{+0.04}_{-0.06}$ & $3.16^{+0.01}_{-0.02}$ & 170 & 0.54 \\
   \parbox[c][17pt][c]{0cm}{} Sculptor & $-1.91^{+0.07}_{-0.06}$ & $3.88^{+0.02}_{-0.02}$ & 252 & 0.00 & & $-1.01^{+0.11}_{-0.11}$ & $3.72^{+0.02}_{-0.02}$ & 243 & 0.00 \\
   \parbox[c][17pt][c]{0cm}{} Leo~II & $-2.43^{+0.16}_{-0.11}$ & $3.72^{+0.13}_{-0.24}$ & 110 & 21 & & $-1.75^{+0.11}_{-0.13}$ & $3.80^{+0.02}_{-0.03}$ & 113 & 0.00 \\
   \parbox[c][17pt][c]{0cm}{} Sextans & $-2.05^{+0.13}_{-0.13}$ & $4.13^{+0.07}_{-0.09}$ & 111 & 0.00 & & $-1.09^{+0.30}_{-0.26}$ & $4.04^{+0.04}_{-0.06}$ & 108 & 0.00
   \enddata
   \label{tab:results}
  \end{deluxetable*}
  \begin{figure*}
   \begin{center}
    \plottwo{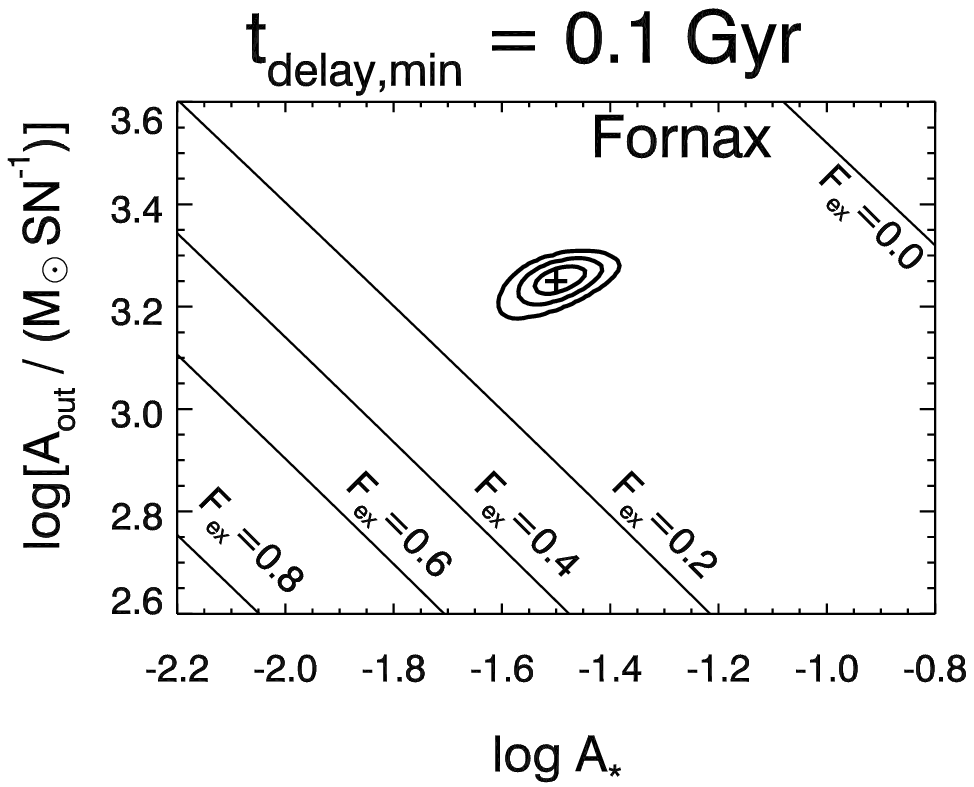}{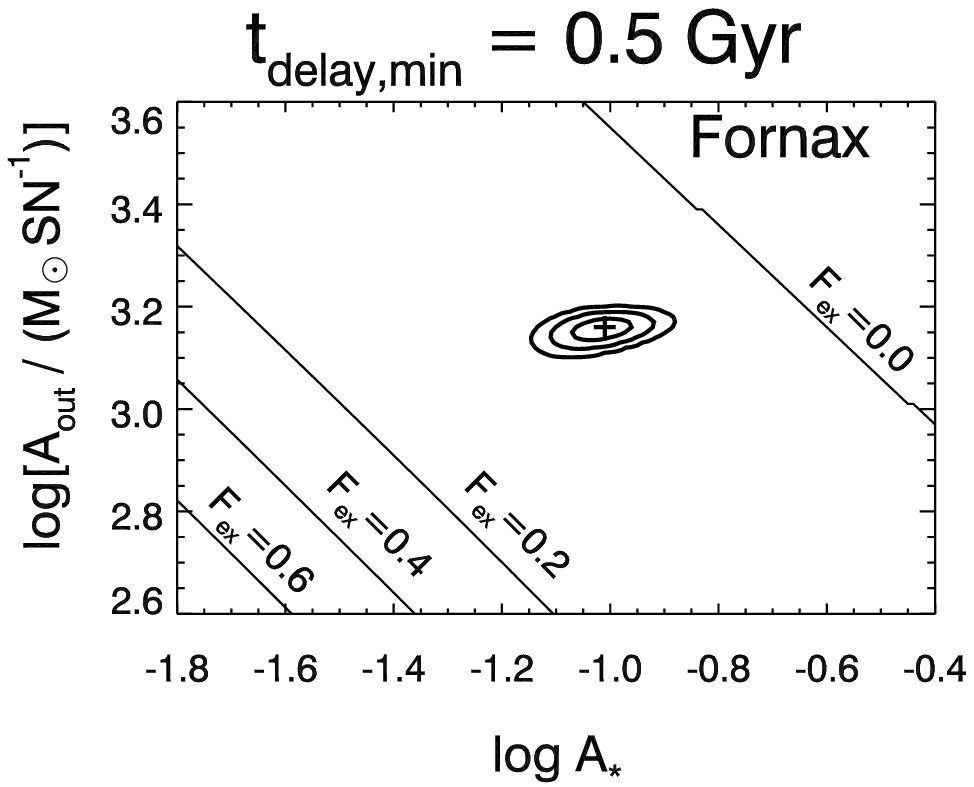}\\
    \plottwo{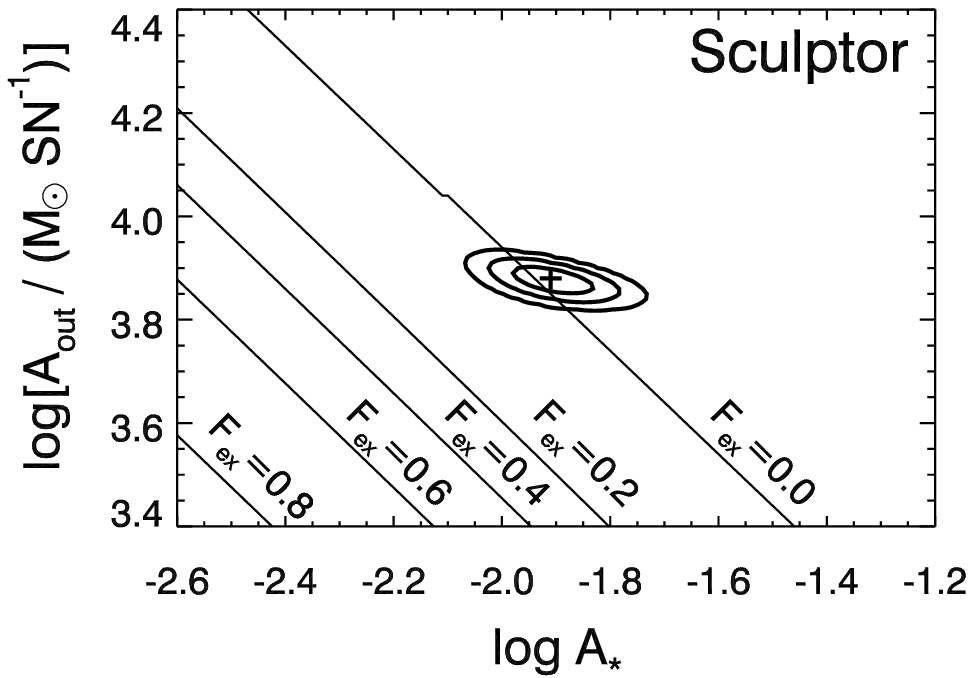}{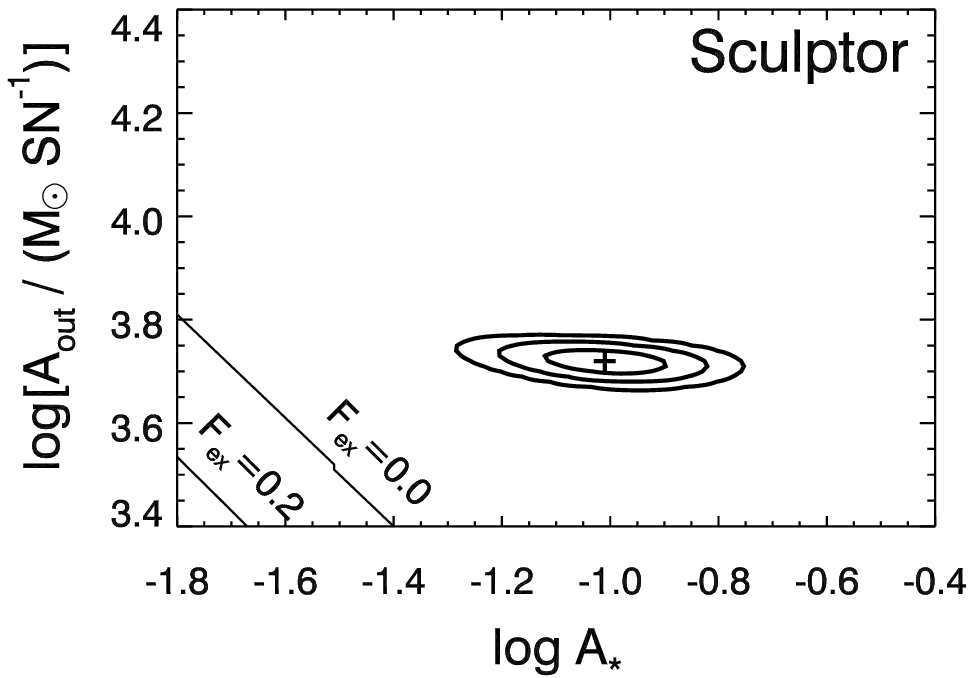}\\
    \plottwo{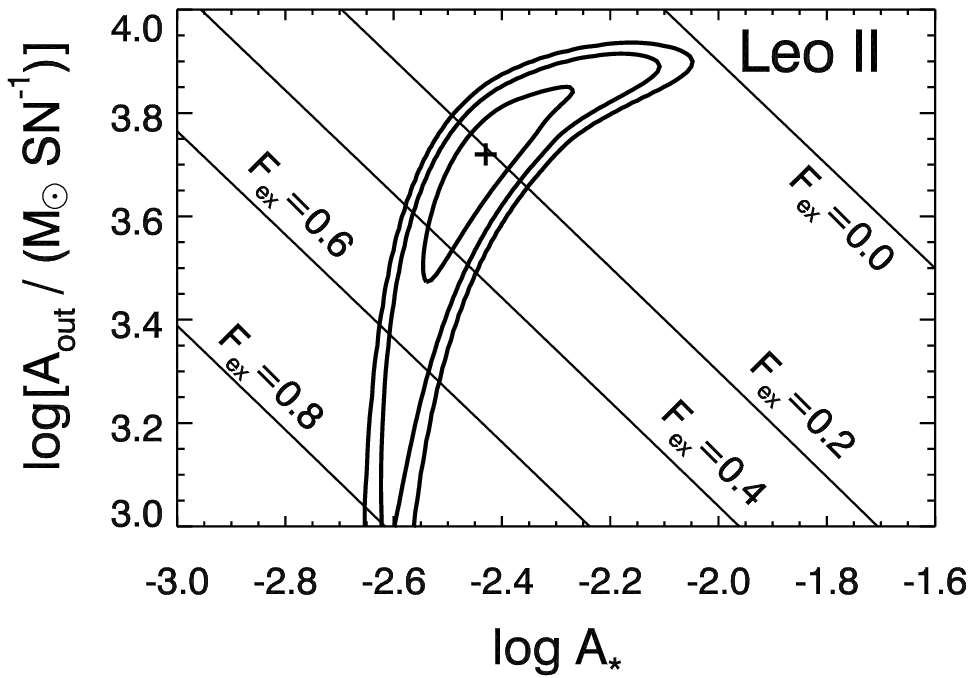}{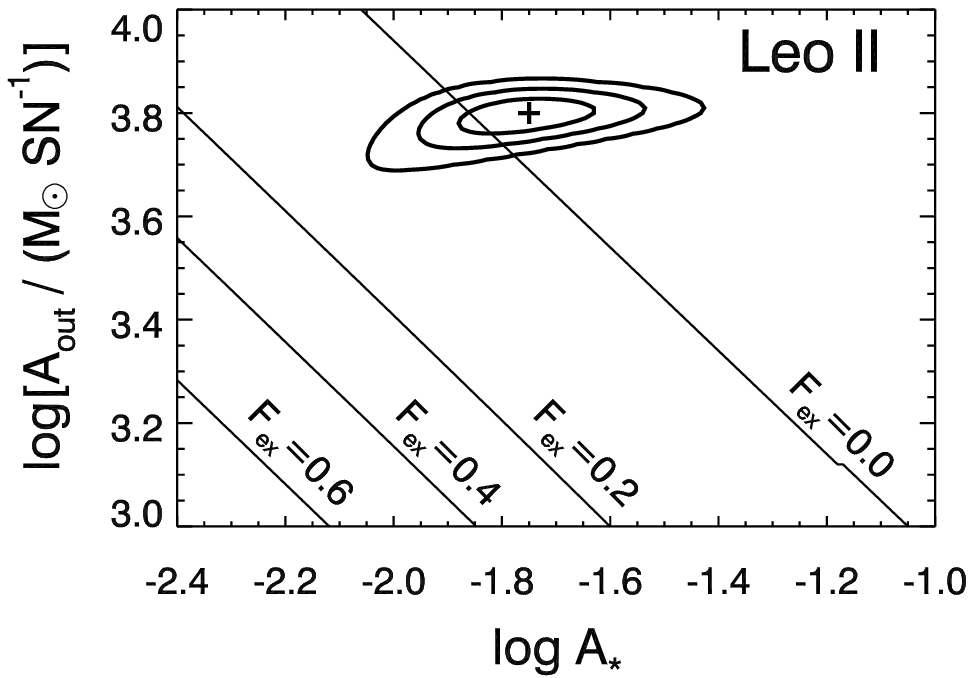}\\
    \plottwo{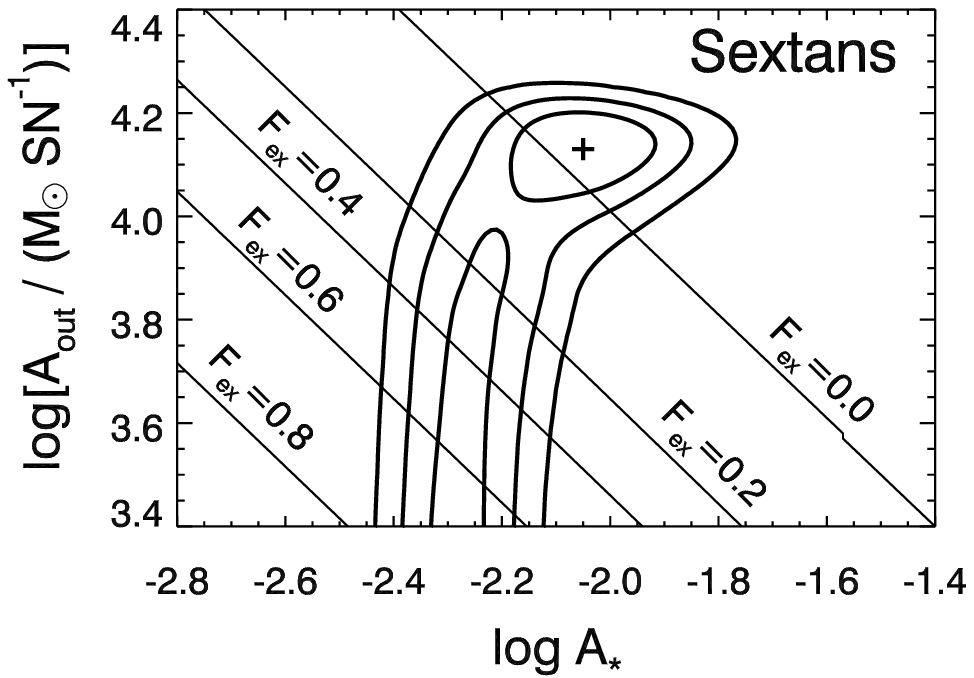}{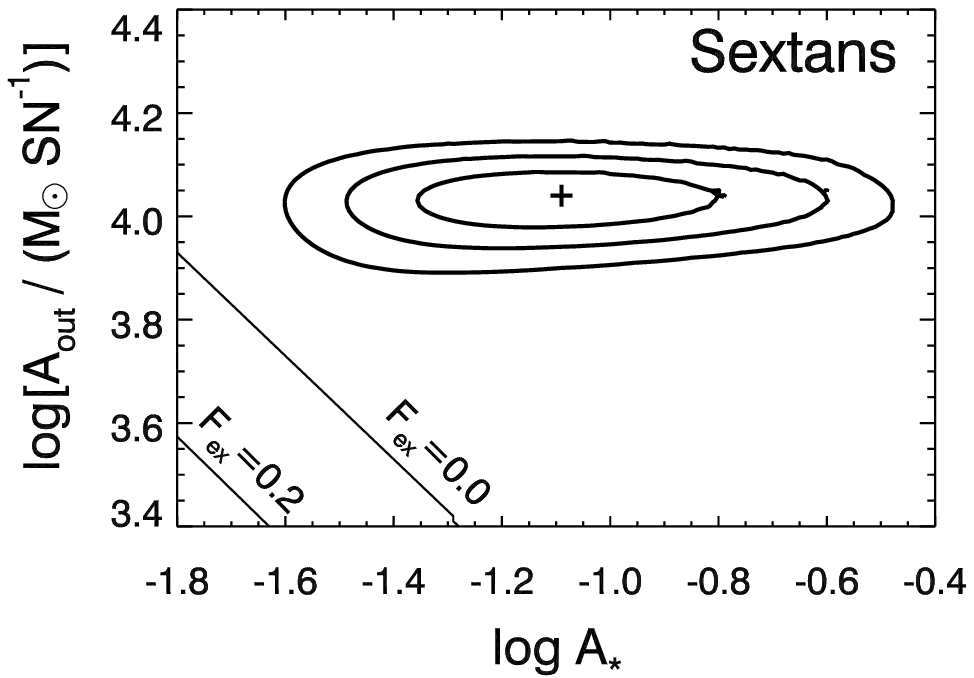}\\
    \caption{Confidence levels for the model parameters, $A_\mathrm{out}$ and $A_\ast$, with $t_\mathrm{delay, min} = 0.1$~Gyr (left) and 0.5~Gyr (right).
	Plus symbols represent the best-fit model parameters, and thick solid contours in each panel indicate $1\sigma$, $2\sigma$, and $3\sigma$ confidence levels.
	Thin solid curves show the contours of $F_\mathrm{ex}$ corresponding to 0.0, 0.2, 0.4, 0.6, and 0.8
    as indicated in the key.\label{fig:contour}}
   \end{center}
  \end{figure*}
  The best-fit model parameters with their $1\sigma$ uncertainties are listed in Table~\ref{tab:results} and shown in Figure~\ref{fig:contour}.
  It is important to note again that these uncertainties do not include the uncertainties in SFHs shown in Figure~\ref{fig:SFH}, as described in Section~\ref{subsubsec:SFH}.
  If the uncertainties in the SFH are included, the derived uncertainties of the best-fit model parameters will be larger.

  As shown in Figure~\ref{fig:contour}, both of the model parameters $A_\ast$ and $A_\mathrm{out}$ are well determined using the observed MDFs except those with $t_\mathrm{delay, min} = 0.1$~Gyr for Leo~II and Sextans, whose contours are apparently elongated.
  Such elongated contours of constant confidence levels are attributed to large $F_\mathrm{ex}$.
  Large $F_\mathrm{ex}$ implies that the extra gas outflow contributes significantly to the outflow from the system rather than the superwind; therefore, the model results depend weakly on $A_\mathrm{out}$.

  For each dSph, the likelihoods of the best-fit models with $t_\mathrm{delay,min} = 0.1$~Gyr and 0.5~Gyr are found to be similar with each other as shown in Table~\ref{tab:results}.
  This result implies that the best-fit value of $t_\mathrm{delay,min}$ for a certain dSph cannot be evaluated by analyzing its MDF alone, while $A_\mathrm{out}$ and $A_\ast$ can be determined by the metallicity at which the MDF has a peak and the significance of the metal-poor tail in the MDF, respectively.
  However, this difficulty for $t_\mathrm{delay,min}$ can be solved through a comparison of the model with observed data in the $\mathrm{[Mg/Fe]}$--$\mathrm{[Fe/H]}$ plane, as shown in the right panels of Figure~\ref{fig:MDF}.
  It is clearly shown that, for all of the sample dSphs, the evolution tracks of the model with $t_\mathrm{delay, min} = 0.5$~Gyr in the plane provide better fits to the observed data than those with $t_\mathrm{delay, min} = 0.1$~Gyr.
  This difference of the evolution track in the $\mathrm{[Mg/Fe]}$--$\mathrm{[Fe/H]}$ plane is originated from the difference between the SNe~Ia and II yields.
  Since the SN Ia yields have smaller abundance ratio of $[\alpha / \mathrm{Fe}]$ than the SN~II yields, $[\alpha / \mathrm{Fe}]$ of the system start to decrease when SN~Ia starts to explode, that is, when the time of $t_\mathrm{delay,min}$ is elapsed since the onset of star formation.

  Through this comparison for our sample dSphs, it is found that $t_\mathrm{delay, min} = 0.1$~Gyr is too short to fit the observed distribution of the stars in the $\mathrm{[Mg/Fe]}$--$\mathrm{[Fe/H]}$ plane.
  However, this result is clearly against the observational estimates for the minimum delay time of SN~Ia, which provides $t_\mathrm{delay, min} \lesssim 0.1$~Gyr almost certainly \citep{Totani_08, Maoz_10}.
  This discrepancy between our model prediction and the observational estimate to $t_\mathrm{delay, min}$ is discussed in Section~\ref{subsec:DiscrepTdelaymin}.

  \subsection{Time Evolution of Mass and Gas Flow of the Best-fit Models}\label{sec:3-2}

  \begin{figure*}
   \begin{center}
    \plotone{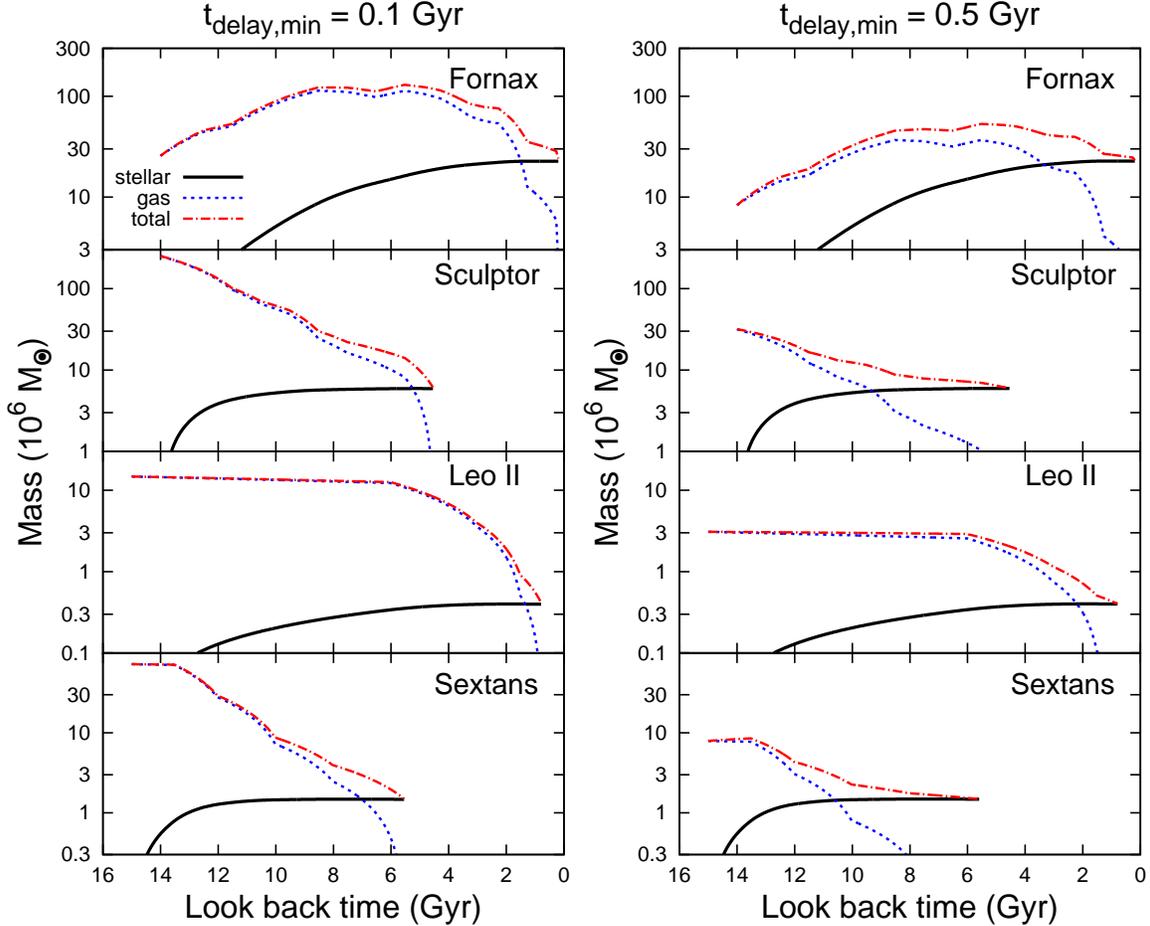}
    \caption{Time evolution of various masses calculated from the best-fit models for our sample dSphs.
	Left and right panels show the best-fit models with $t_\mathrm{delay,min} = 0.1$~Gyr and 0.5~Gyr, respectively.
	The black solid and blue dotted curves represent the stellar masses and interstellar gas masses, respectively, while the red dot-dashed curves show their total masses.\label{fig:Mass}}
   \end{center}
  \end{figure*}
  We show the time evolution of stellar mass $M_\ast (t)$ and total mass of interstellar gas $M_\mathrm{gas} (t)$ as well as the total mass of the system $M_\ast (t) + M_\mathrm{gas} (t)$ calculated from the best-fit models for our sample dSphs in Figure~\ref{fig:Mass}.
  Note that, since the observed SFH is adopted as the input in our model, the time evolution of stellar mass is completely independent of model parameters; therefore, both of the shapes and absolute values of $M_\ast (t)$ for a dSph calculated from the best-fit models with $t_\mathrm{delay, min} = 0.1$~Gyr and 0.5~Gyr are identical with each other.
  Moreover, since gas mass at a given time is assumed to be linearly proportional to the SFR at the time, $\Psi (t)$, according to Equation~(\ref{eq:mass}), the time evolution of gas mass for a certain dSph has exactly same shape as that of its SFR by definition.
  Therefore, the shape of $M_\mathrm{gas} (t)$ is independent of the model parameters as shown in Figure~\ref{fig:Mass}.  However, the ratio of $M_\mathrm{gas} (t) / \Psi (t)$ is controlled by one of the model parameters $A_\ast$ in the sense that smaller $A_\ast$ results in larger gas mass for a given SFR.
  For the dSphs for which $A_\ast$ of the best-fit models are very low (i.e., $\log{A_\ast} \lesssim -2$), that is, Sculptor, Leo~II, and Sextans with with $t_\mathrm{delay,min} = 0.1$~Gyr, gas mass dominates the total mass of the dSph in almost entire time during the calculation.

  \begin{figure*}
   \begin{center}
    \plotone{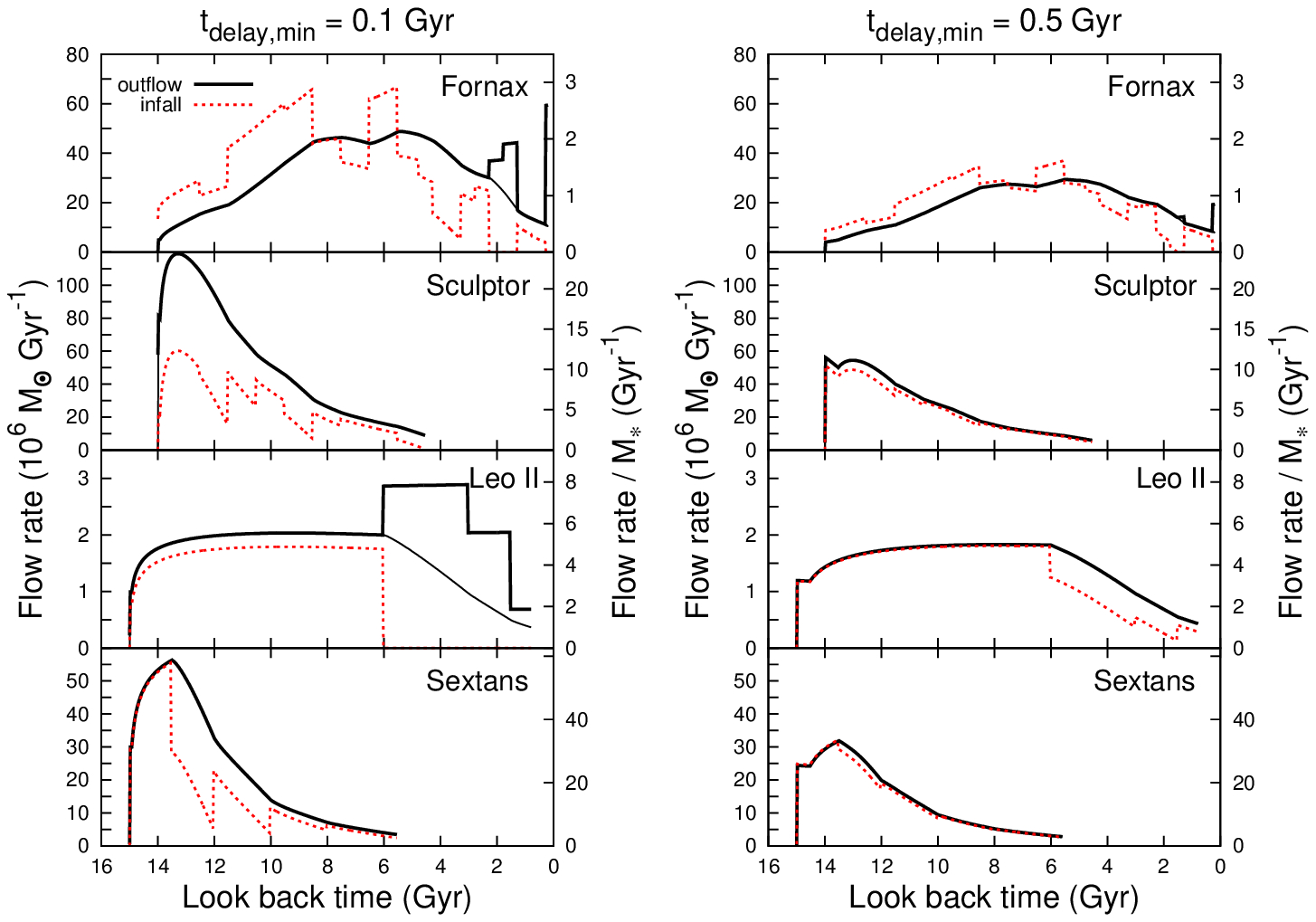}
    \caption{Same as Figure~\ref{fig:Mass} but for the time evolution of the outflow and infall rates.
	The vertical axis on the right side indicates the outflow and infall rates per present stellar mass of each dSph.
	The present stellar masses of the Fornax, Sculptor, Leo II, and Sextans are $M_\ast = 2.3\times10^7$, $4.9\times10^6$, $3.7\times10^5$, and $9.6\times10^5~M_\odot$, respectively, which are estimated by their SFHs.
	The black solid and red dotted curves represent the outflow and infall rates, respectively.
	For the outflow rates, the thin and thick black curves present the original outflow rates of $\dot{M}_{\mathrm{out}} (t)$, which are evaluated via Equation~(\ref{eq-Mout}), and the net outflow rates of $\dot{M}_{\mathrm{out}} (t) + \dot{M}_\mathrm{out}^\mathrm{ex} (t)$, respectively.\label{fig:Flow}}
   \end{center}
  \end{figure*}
  The time evolution of the outflow and infall rates of the best-fit models are shown in Figure~\ref{fig:Flow}.
  It is found that, while outflow rate evolves smoothly, the shape of the time evolution of infall rate is like a step function except for Leo~II.
  This step-function like shape of infall rate is originated from the prescription for infall rate in our model; that is, the infall rate is adjusted to satisfy the condition of Equation~(\ref{eq:sec2-2}) for a given SFH.
  Thus, the infall rate can suddenly change at the time when the slope of SFR changes (see Figure~\ref{fig:SFH}).
  In this sense, such step-function like evolution of infall rate is rather artificial result caused by using the observationally estimated SFH with coarse time resolution.
  On the other hand, since the outflow rate is determined by the number of SN~Ia and SN~II as described in Equation (\ref{eq-Mout}) and hence the SFR, the time evolution of outflow rate looks similar to the SFH and smoother than infall rate.

  \begin{figure}
   \begin{center}
    \plotone{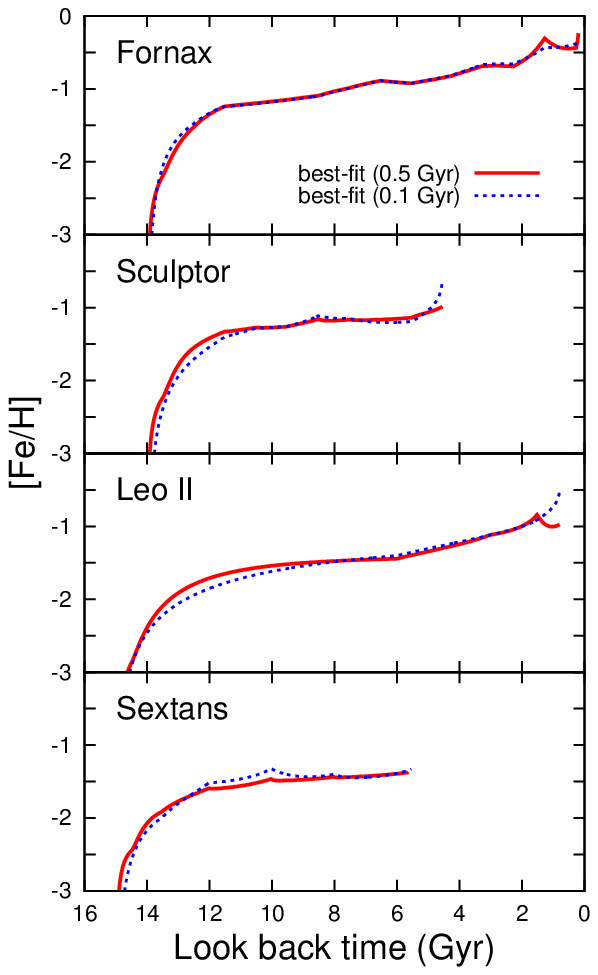}
    \caption{The AMRs calculated from the best-fit models for our sample dSphs.
	The best-fit models with $t_\mathrm{delay, min} = 0.1$~Gyr and 0.5~Gyr are shown as blue dotted and red solid curves, respectively.\label{fig:AMR}}
   \end{center}
  \end{figure}
  For the best-fit models for Leo~II with $t_\mathrm{delay,min} = 0.1$~Gyr and Fornax with both $t_\mathrm{delay,min}$, bumps are also seen in the outflow rates at the late stages of evolution: that is, the look back time of $\sim 1$--6~Gyr for Leo~II and $\sim 2$~Gyr and $\sim 0$~Gyr for Fornax.
  These bumps are originated from the contributions of the extra outflow gas masses to avoid negative infall rate explained in Section~\ref{sec:infall}.
  The fractions of extra outflow gas masses $F_{\rm ex}$ of these models are $F_{\rm ex}=$ 4.4\%, 0.54\%, and 21\% for the best-fit models for Fornax with $t_{\rm delay,min} = 0.1~{\rm Gyr}$ and 0.5~Gyr and for Leo~II with $t_{\rm delay,min} = 0.1~{\rm Gyr}$, respectively.
  This artificial prescription for the extra outflow tends to appear in the late stages of evolution, where the increasing rate of metallicity of the system is smaller than that in the early stage as shown in Figure~\ref{fig:AMR}.
  Therefore, the effect of this extra gas outflow to the MDF is small.

  The gas flow rates of the models with lower $A_\ast$ are larger since the gas masses are larger.
  Total outflow gas masses are $\sim 10$--100 times larger than the present stellar masses.
  It is worth noting that the outflow and infall rates are comparable.
  This result implies that a large amount of infalling gas, which is comparable to outflowing gas associated with star formation, is inevitably required to reproduce both low average metallicity and relatively long SFH of dSph simultaneously.
  Specifically, in order to explain the observed low metallicities of dSphs, a large amount of metals is required to be blown away by superwinds.
  However, since a large amount of gas outflow terminates the star formation in dSphs rapidly, a comparable gas infall is necessary to reproduce relatively long SFHs estimated observationally for dSphs.
  We will provide a discussion for the chemical evolution of dSph and comparison with the results of previous works in Section~\ref{sec:4-4}.

 \section{Discussion}\label{sec:discussion}

  \subsection{The Dependence of Chemical Evolution on the Model Parameters}\label{sec:4-1}

  \begin{figure}
   \begin{center}
    \plotone{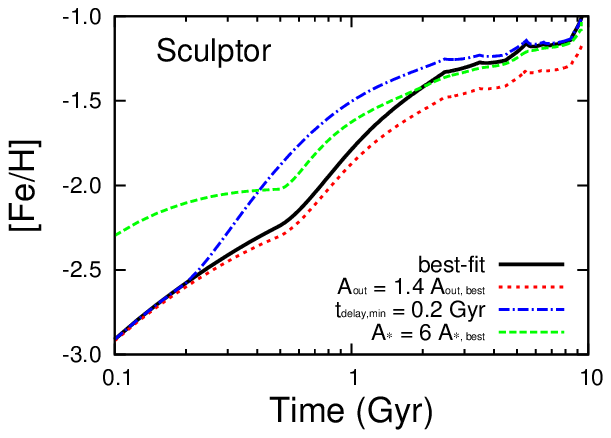}
    \caption{Comparison of the AMRs calculated from the models with various sets of the model parameters of $A_\mathrm{out}$, $t_\mathrm{delay,min}$, and $A_\ast$ including the best-fit model parameters for Sculptor.
	While the black curve represents the AMR of the best-fit model with $t_\mathrm{delay,min} = 0.5$~Gyr, the other curves show the AMRs of the models with the parameters one of which is changed from the best-fit one.
	For the red dotted, blue dot-dashed, and green dashed curves, the parameters changed from the best-fit one are $A_\mathrm{out}$, $t_\mathrm{delay,min}$, and $A_\ast$, respectively, as indicated in the key.\label{fig:AMR_compare}}
   \end{center}
  \end{figure}
  In Section~\ref{sec:best-fit}, we show that the observed MDFs of our sample dSphs are reasonably reproduced by our best-fit models.
  Here we investigate the detailed dependence of MDF on our model parameters of $A_\ast$, $A_{\rm out}$, and $t_\mathrm{delay, min}$.
  Considering that the MDF is roughly determined by the SFH and the AMR as
  \begin{equation}
   \frac{dN_\mathrm{stars}}{d\mathrm{[Fe/H]}} \propto \frac{d\Psi}{dt} \bigg/ \frac{d\mathrm{[Fe/H]}}{dt},
  \end{equation}
  and that the SFH is fixed in our model, the MDF will depend only on the AMR.
  Therefore, through investigating the effects of the model parameters of $A_\ast$, $A_\mathrm{out}$, and $t_\mathrm{delay,min}$ on the AMR which are shown in Figure~\ref{fig:AMR_compare}, we can examine their effects on the MDF.
  Note that, while the illustrated models in Figure~\ref{fig:AMR_compare} are only the cases for Sculptor, the AMRs for the other dSphs show the similar dependence on the model parameters.

  As shown in Figure~\ref{fig:AMR_compare}, compared with the AMR of the best-fit model with $t_\mathrm{delay,min} = 0.5$~Gyr (black solid curve), the AMR of the model with larger $A_\mathrm{out}$ (red dotted curve) results in a slower metal evolution and its deviation from the best-fit model increases monotonically with time.
  This is because, for a given SFR, a larger $A_\mathrm{out}$ leads to a more amount of metal elements expelled from the system as a more intense outflow, resulting in a smaller growth rate of metallicity.
  Since the deviation from the best-fit model seems to be evident throughout the late stages of chemical evolution, $A_\mathrm{out}$ affects the mean metallicity of the system; larger $A_\mathrm{out}$ leads to smaller mean metallicity.  In fact, the best-fit models for our sample dSphs show this trend (see Table~\ref{tab:results} and Figure~\ref{fig:MDF}).
  In other words, our model predicts that the observed low metallicities of dSphs are the consequence of intense gas outflows accompanied by large amount of metal.

  On the other hand, the effect of changing $t_\mathrm{delay,min}$ from the value in the best-fit model of $t_\mathrm{delay,min} = 0.5$~Gyr into 0.2~Gyr appears in the early phase of chemical evolution as shown by the blue dot-dashed curve in Figure~\ref{fig:AMR_compare}.
  Since the SN~Ia supplies much abundant Fe to ISM compared to SN~II and AGB, metallicity of the system increases rapidly just after the SN~Ia starts to explode; that is, the minimum delay time $t_\mathrm{delay, min}$ is elapsed since the onset of star formation.
  Therefore, the AMRs with different $t_\mathrm{delay, min}$ are clearly different in the early stages of chemical evolution.
  However, the growth rate of metallicity decreases with time and hence the difference between the models with different $t_\mathrm{delay,min}$ becomes negligible in the late phases of chemical evolution.
  Therefore, the effect of $t_\mathrm{delay,min}$ is important in the early phase of chemical evolution, in particular around $t \sim t_\mathrm{delay,min}$.

  The effect of changing $A_\ast$ appears also in the early phase of chemical evolution as shown by the green dashed curve in Figure~\ref{fig:AMR_compare}.
  Since a higher $A_\ast$ means that the system has smaller gas mass for a given SFR according to Equation~(\ref{eq:mass}), the interstellar gas of the system is easily enriched by the yields from the evolved stars, which are fixed from the past SFH.
  Therefore, the model with a higher $A_\ast$ results in a more rapid chemical enrichment in the early phase.
  However, the growth rate of metallicity decreases with time similar to the model with changing $t_\mathrm{delay,min}$.
  As a result, changing $A_\ast$ affects the metal evolution mainly in the early phase.

  Specifically, a lower $A_\ast$ results in a more number of metal-poor stars, producing a broader metal-poor tail in the MDF.
  In other words, the dSph whose MDF has a broad metal-poor tail is expected to prefer a low $A_\ast$.  
  However, the best-fit parameters of $A_\ast$ show no relation with the broadness of the metal-poor tails in the MDFs against the expectation from our model; while Sculptor has a broader metal-poor tail than Fornax and Leo~II, the best-fit parameter of $A_\ast$ for Sculptor is similar to that for Fornax and larger than that for Leo~II (see Table~\ref{tab:results}).
  This contradiction comes from the fact that the effect of $A_\ast$ on the metal-poor tail in MDF is not primary but secondary.
  The broadness of the metal-poor tail in MDF is primarily determined by the SFH, or more specifically, the fraction of the stars formed in early phase of chemical evolution.
  As shown in Figure~\ref{fig:SFH}, the fraction of the stars formed in early phase is large in Sculptor but small in Fornax and Leo~II.
  When we calculate the chemical evolution for Fornax, Sculptor, and Leo~II with same $A_\ast$, the metal-poor tails in the resultant MDFs would be broader in Sculptor and narrower in Fornax and Leo~II.
  In other words, the value of $A_\ast$ is not determined only by the shape of the metal-poor tail in MDF but also by the shape of SFH.

  Moreover, the similarity between the effects of changing $t_\mathrm{delay,min}$ and $A_\ast$ on AMR leads to the degeneracy between these parameters in our chemical evolution model.
  As shown in Figure~\ref{fig:MDF} and Table~\ref{tab:results}, the observed MDFs can be reproduced regardless of $t_\mathrm{delay,min}$ since the model MDF with short $t_\mathrm{delay,min}$ and low $A_\ast$ is similar to the model with long $t_\mathrm{delay,min}$ and high $A_\ast$.
  This degeneracy can be resolved by comparing the model results with the observed $\mathrm{[Mg/Fe]}$--$\mathrm{[Fe/H]}$ diagrams simultaneously since different $t_\mathrm{delay,min}$ results in different abundance ratios of $\alpha$ elements as described in Section~\ref{subsec:Param+Errors}.
  Therefore, both the observed MDF and abundance ratio are required to determine $t_\mathrm{delay,min}$ and $A_\ast$ accurately.

  In summary, it is difficult to understand the chemical evolution of dSphs only by analyzing the MDF and abundance ratio.
  It is important to examine MDF, abundance ratio, and SFH simultaneously for investigating the chemical evolution of dSphs.

  \subsection{The Effects of the Difference in Observed Area on the SFH and MDF}\label{sec:4-2}

  As already stated in Section~\ref{sec:likelihood}, the observed area for a dSph in which its MDF is evaluated is typically different from that to obtain SFH.
  In our model, the difference in the observed area is completely neglected since dSph is assumed to be chemically homogeneous.
  However, photometric observations for dSphs have reported that some dSphs have radial gradient with age and metallicity; young stars are more concentrated than old stars and metal-rich stars are more concentrated than metal-poor stars \citep{de_Boer_12a, de_Boer_12b, Lee_09}.
  Therefore, the dSphs are not chemically homogeneous and the difference in the observed area for the SFH and MDF may affect our results.

  \begin{figure*}
   \begin{center}
    \plotone{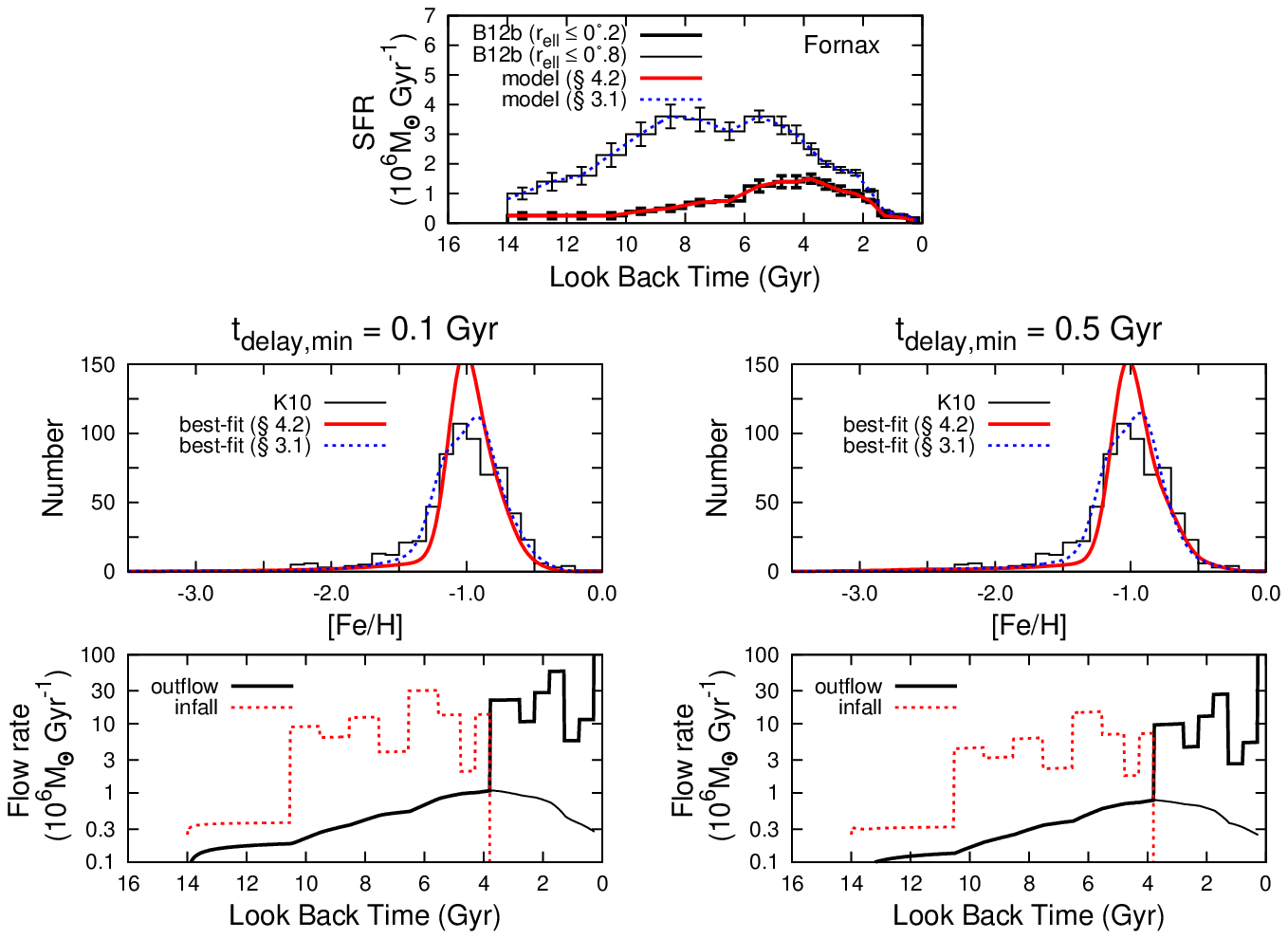}
    \caption{(Top) The SFH of the central region ($r_\mathrm{ell} < 0\fdg 2$) of Fornax obtained by \citet{de_Boer_12b} and the SFH used in our model calculation represented by the thick histogram and red solid curve, respectively.
	For reference, the SFHs shown in Figure~\ref{fig:SFH} are also presented as thin histogram and blue dotted curve.
	(Middle) The observed and best-fit model MDFs for the SFH of the central region of Fornax shown as histograms and red solid curves, respectively.
	Note that the observed MDF shown as histogram is the same as that shown in the top-left panel of Figure~\ref{fig:MDF}.
	The best-fit models with the parameters listed in Table~\ref{tab:results} are also plotted by blue dotted curves, for reference.
	(Bottom) Same as Figure~\ref{fig:Flow} but for the time evolution of the outflow and infall rates in the best-fit models for the SFH and MDF in the central region of Fornax.
	In the middle and bottom, left and right panels show the best-fit models with $t_\mathrm{delay,min} = 0.1$~Gyr and 0.5~Gyr, respectively.\label{fig:fornax_inner}}
   \end{center}
  \end{figure*}
  \begin{figure*}
   \begin{center}
    \plotone{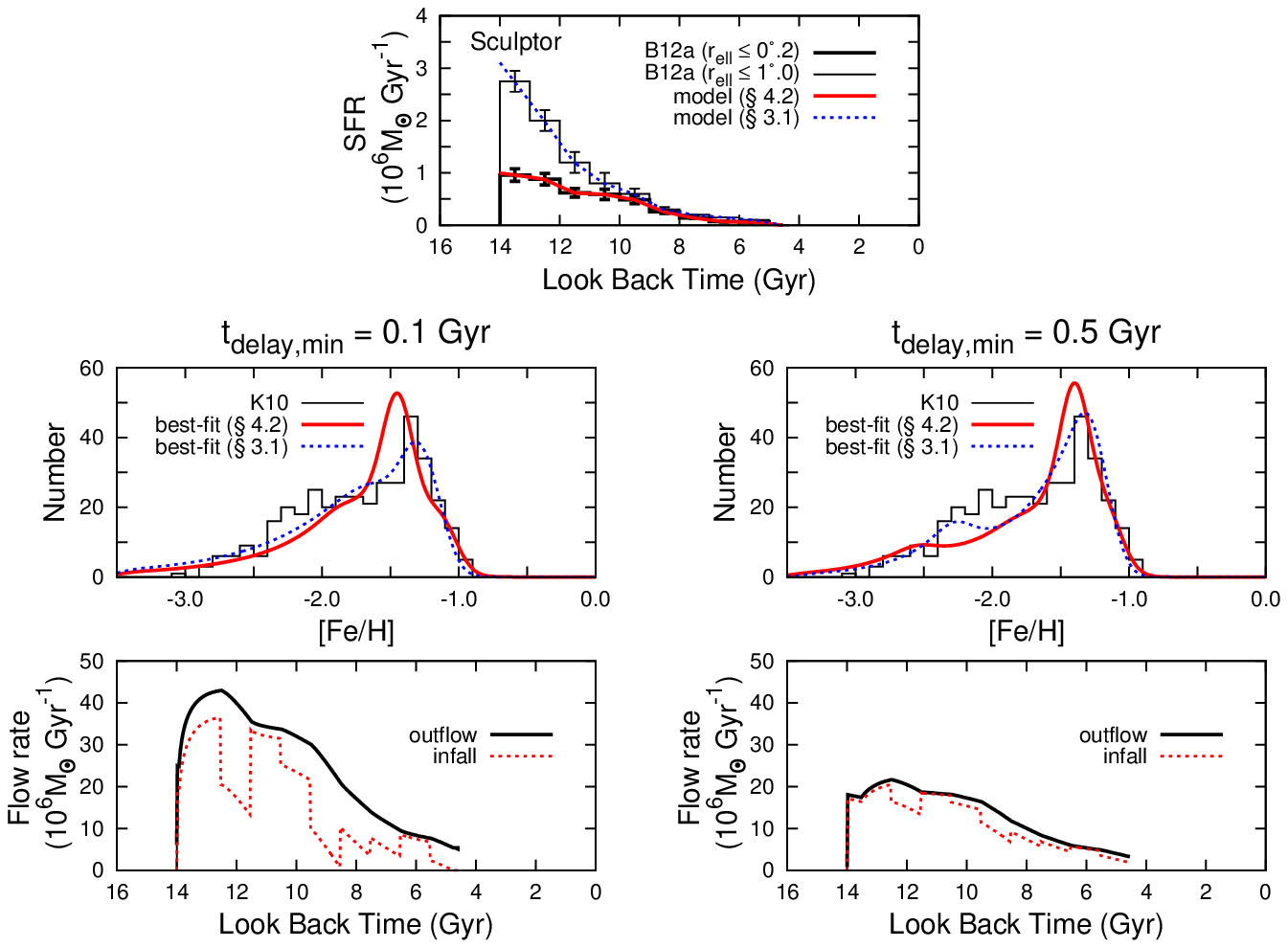}
    \caption{Same as Figure~\ref{fig:fornax_inner} but for Sculptor.\label{fig:sculptor_inner}}
   \end{center}
  \end{figure*}
  In order to investigate the effect of this difference on the model results, we calculate the chemical evolution for Fornax and Sculptor by using their SFHs and MDFs evaluated from the stars in the same area, that is, the central region ($r_\mathrm{ell} < 0\fdg 2$) obtained by \citet{de_Boer_12a, de_Boer_12b}, where $r_\mathrm{ell}$ is the elliptical radius of the dSphs used in \citet{de_Boer_12a, de_Boer_12b}.
  As shown in Figures~\ref{fig:fornax_inner} and \ref{fig:sculptor_inner}, while the number of metal-poor stars in the observed MDFs is slightly underestimated, our model can reproduce the observed MDFs for Fornax and Sculptor with almost same parameters except $A_\mathrm{out}$ for Fornax\footnote{The best-fit parameters of $A_\ast$ with $t_\mathrm{delay,min} = 0.1$~Gyr $(0.5~\mathrm{Gyr})$ for Fornax and Sculptor are $\log{A_\ast} = -1.76^{+0.04}_{-0.01}$ $(-1.43^{+0.02}_{-0.01})$ and $-1.95^{+0.07}_{-0.06}$ $(-1.36^{+0.11}_{-0.08})$, respectively. The best-fit parameters of $A_\mathrm{out}$ with $t_\mathrm{delay,min} = 0.1$~Gyr $(0.5~\mathrm{Gyr})$ for Sculptor are $\log{[A_\mathrm{out} / (M_{\odot}~\mathrm{SN}^{-1})]} = 3.86^{+0.03}_{-0.04}$ $(3.72^{+0.02}_{-0.02})$. On the other hand, since $F_\mathrm{ex}$ of the best-fit models for Fornax are large, the best-fit parameters of $A_\mathrm{out}$ are not determined well. The $3\sigma$ upper limit of $A_\mathrm{out}$ is $\log [A_\mathrm{out} / (M_{\odot}~\mathrm{SN}^{-1})] = 3.02$ $(2.73)$ with $t_\mathrm{delay,min} = 0.1$~Gyr $(0.5~\mathrm{Gyr})$.}, which requires for a large amount of extra gas outflow (i.e., $F_\mathrm{ex} > 0.8$) in the last several Gyrs.

  \begin{figure*}
   \begin{center}
    \plotone{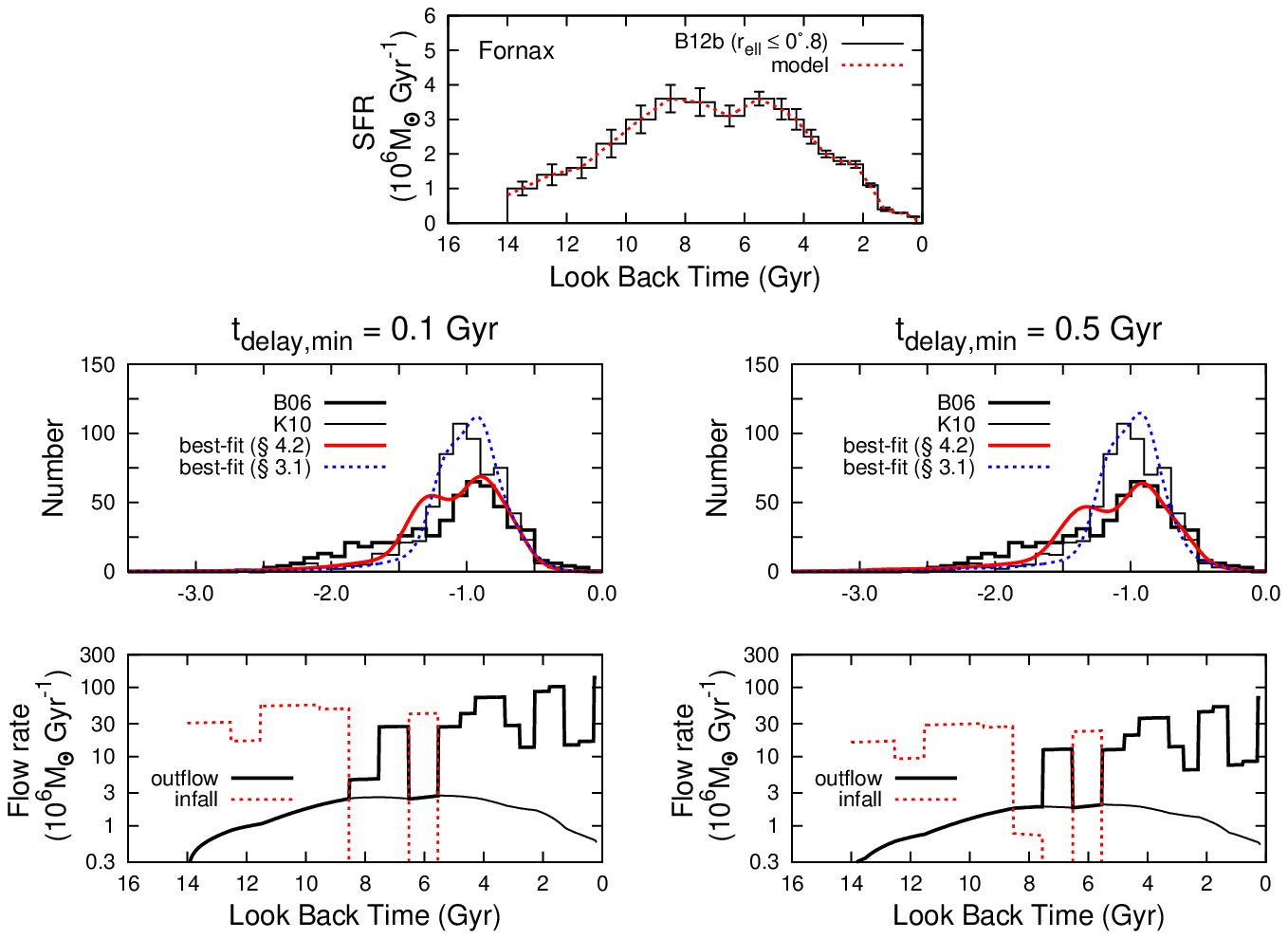}
    \caption{Same as Figure~\ref{fig:fornax_inner} but for the observed data and best-fit models for the SFH and MDF in a wide area of Fornax.
	The MDF of a wide area obtained by \citet{Battaglia_06} is shown as thick histogram in the middle panels, while the observed SFH and MDF shown as thin histogram in the top and middle panels are the same as those plotted in the top panel of Figure~\ref{fig:SFH} and top-left panel of Figure~\ref{fig:MDF}, respectively.\label{fig:fornax_Battaglia06}}
   \end{center}
  \end{figure*}
  We also calculate the chemical evolution for Fornax by using the MDF evaluated from the 562 stars in a wide area ($\sim 1\fdg 0 \times 1\fdg 5$) of Fornax \citep{Battaglia_06}, which is comparable with the observed area for SFH by \citet{de_Boer_12b}.
  Although the observed MDF in the wide area shows a more number of metal-poor stars than the MDF shown in Figure~\ref{fig:MDF}, our model cannot reproduce it as shown in Figure~\ref{fig:fornax_Battaglia06}.
  Moreover, a significant amount of extra outflow (i.e., $F_\mathrm{ex} > 0.8$) is required for the last $\sim 8$~Gyrs as shown in the bottom panels of Figure~\ref{fig:fornax_Battaglia06}.

  The results of the above comparison indicate that our model for chemical evolution cannot explain both the SFH and MDF simultaneously in the central and/or the whole regions of Fornax.
  This difficulty can be relaxed if the SFE (i.e., $A_\ast$ with the units of $\mathrm{Gyr^{-1}}$) is adopted to be time variant, which is suggested from observational studies (e.g., \citealt{Combes_13}).
  Since the SFE controls the growth rate of metallicity in the early phase as described in Section~\ref{sec:4-1}, the model with a time variable SFE can form a more number of metal-poor stars.

  A multi-zone model is also expected to resolve the difficulty in our model for the number of metal-poor stars.
  Some studies presented a signature of multi-component structure in dSphs.
  It is reasonable that the metal-poor population in a dSph has experienced different chemical evolution from the metal-rich population.
  Hence, it is important to gather many observational results of dSphs and construct a more realistic multi-zone chemical evolution model.

  \subsection{Comparison of the Best-Fit Parameters with Those of K11}\label{sec:K11}

  As described in Section~\ref{sec:assumption}, our model is basically the same as that of K11.
  The best-fit parameters in K11 for our sample dSphs are listed in Table~\ref{tab:cases}.
  Since the unit of $A_\ast$ in our model is different from that in the K11 model ($A_\ast^\mathrm{K11}$), we show a converted values of $A_\ast$ from $A_\ast^\mathrm{K11}$ in Table~\ref{tab:cases} by $A_\ast = A_\ast^\mathrm{K11} / (10^6~M_\odot~\mathrm{Gyr}^{-1}$).
  Then we can directly compare the best-fit parameters derived from the K11 model with ours.

  It is found that, while the values of $A_\mathrm{out}$ in K11 are similar to those in our best-fit models, $A_\ast$ of our best-fit models with $t_\mathrm{delay,min} = 0.1$~Gyr are much smaller than those of K11.
  To investigate the origin of this difference, we explore the best-fit models for the following four cases: (Case~1) the model with the fixed SFH, $\alpha$, and $A_\mathrm{out}$ which are the same as those in K11 and with only one adjustable parameter of $A_\ast$, (Case~2) the model with the same SFH and $\alpha$ as those in K11 and with adjustable parameters of $A_\ast$ and $A_\mathrm{out}$, (Case~3) the model with the same SFH as that in K11 and with three adjustable parameters $\alpha$ (fixed to unity), $A_\ast$, and $A_\mathrm{out}$, and (Case~4) the model with the same shape of the SFH as K11 but its timescale is extended to fit the timescale of our SFH.
  The resultant best-fit parameters for the four cases are listed in Table~\ref{tab:cases}.
  \begin{deluxetable*}{llcccccc}
   \tablecolumns{8}
   \tablecaption{The best-fit model parameters with $t_\mathrm{delay,min} = 0.1$~Gyr of K11, our model, and comparison models for our sample dSphs}
   \tablehead{
   \colhead{} & \colhead{} & \colhead{} & \multicolumn{4}{c}{Comparison model} & \colhead{} \\ \cline{4-7}
   \colhead{dSph} & \colhead{Parameter} & \colhead{K11} & \colhead{Case~1} & \colhead{Case~2} & \colhead{Case~3} & \colhead{Case~4} & \colhead{Our model}
   }
   \startdata
   \parbox[c][17pt][c]{0cm}{} & $\alpha$ & $0.98^{+0.15}_{-0.04}$ & 0.98 (fixed) & 0.98 (fixed) & 1.0 (fixed) & 1.0 (fixed) & 1.0 (fixed) \\
   \parbox[c][17pt][c]{0cm}{} Fornax & $\log{A_\ast}$ & $0.70^{+0.16}_{-0.10}$ & $1.14^{+0.06}_{-0.12}$ & $1.14^{+0.11}_{-0.18}$ & $1.14^{+0.12}_{-0.19}$ & $-1.30^{+0.04}_{-0.04}$ & $-1.50^{+0.05}_{-0.03}$ \\
   \parbox[c][17pt][c]{0cm}{} & $\log[A_\mathrm{out} / (M_{\odot}~\mathrm{SN}^{-1})]$ & $3.18^{+0.01}_{-0.02}$ & 3.18 (fixed) & $3.16^{+0.01}_{-0.01}$ & $3.16^{+0.01}_{-0.01}$ & $3.32^{+0.01}_{-0.02}$ & $3.25^{+0.02}_{-0.02}$ \\ \hline
   \parbox[c][17pt][c]{0cm}{} & $\alpha$ & $0.83^{+0.14}_{-0.08}$ & 0.83 (fixed) & 0.83 (fixed) & 1.0 (fixed) & 1.0 (fixed) & 1.0 (fixed) \\
   \parbox[c][17pt][c]{0cm}{} Sculptor & $\log{A_\ast}$ & $-0.33^{+0.08}_{-0.13}$ & $-0.32^{+0.02}_{-0.02}$ & $-0.30^{+0.05}_{-0.04}$ & $-0.41^{+0.06}_{-0.06}$ & $-2.21^{+0.02}_{-0.03}$ & $-1.91^{+0.07}_{-0.06}$ \\
   \parbox[c][17pt][c]{0cm}{} & $\log[A_\mathrm{out} / (M_{\odot}~\mathrm{SN}^{-1})]$ & $3.73^{+0.01}_{-0.01}$ & 3.73 (fixed) & $3.70^{+0.02}_{-0.03}$ & $3.68^{+0.02}_{-0.03}$ & $2.95^{+0.45}_{-0.95}$ & $3.88^{+0.02}_{-0.02}$ \\ \hline
   \parbox[c][17pt][c]{0cm}{} & $\alpha$ & $0.66^{+0.17}_{-0.40}$ & 0.66 (fixed) & 0.66 (fixed) & 1.0 (fixed) & 1.0 (fixed) & 1.0 (fixed) \\
   \parbox[c][17pt][c]{0cm}{} Leo~II & $\log{A_\ast}$ & $-0.37^{+0.50}_{-0.11}$ & $-0.26^{+0.02}_{-0.03}$ & $-0.25^{+0.06}_{-0.06}$ & $-0.48^{+0.12}_{-0.10}$ & $-2.08^{+0.07}_{-0.07}$ & $-2.43^{+0.16}_{-0.11}$ \\
   \parbox[c][17pt][c]{0cm}{} & $\log[A_\mathrm{out} / (M_{\odot}~\mathrm{SN}^{-1})]$ & $3.82^{+0.02}_{-0.02}$ & 3.82 (fixed) & $3.80^{+0.02}_{-0.02}$ & $3.78^{+0.03}_{-0.02}$ & $3.91^{+0.03}_{-0.03}$ & $3.72^{+0.13}_{-0.24}$ \\ \hline
   \parbox[c][17pt][c]{0cm}{} & $\alpha$ & $0.50^{+0.20}_{-0.25}$ & 0.50 (fixed) & 0.50 (fixed) & 1.0 (fixed) & 1.0 (fixed) & 1.0 (fixed) \\
   \parbox[c][17pt][c]{0cm}{} Sextans & $\log{A_\ast}$ & $-0.28^{+0.27}_{-0.18}$ & $-0.29^{+0.03}_{-0.03}$ & $-0.30^{+0.08}_{-0.06}$ & $-0.74^{+0.14}_{-0.13}$ & $-2.47^{+0.14}_{-0.12}$ & $-2.05^{+0.13}_{-0.13}$ \\
   \parbox[c][17pt][c]{0cm}{} & $\log[A_\mathrm{out} / (M_{\odot}~\mathrm{SN}^{-1})]$ & $3.98^{+0.04}_{-0.03}$ & 3.98 (fixed) & $3.91^{+0.09}_{-0.11}$ & $3.91^{+0.07}_{-0.11}$ & $3.97^{+0.15}_{-0.64}$ & $4.13^{+0.07}_{-0.09}$
   \enddata
   \label{tab:cases}
  \end{deluxetable*}

  In the Case~1 model, in which only one parameter of $A_\ast$ can be used to fit the observed MDF, the best-fit values of $A_\ast$ for our sample dSphs are similar to those in K11 as shown in Table~\ref{tab:cases}.
  This result indicates that, if the same SFH and parameters are adopted, the resultant MDF from the K11 and our models are similar with each other.
  Although the best-fit $A_\ast$ is significantly different from that in K11 for Fornax, the discrepancy of $A_\ast$ between the K11 and our model is attributed to the different prescription for gas infall rate.
  Since the infall rate at $t = 0$ is adopted to be exactly equal to zero in the K11 model, high $A_\ast$ results in rapid consumption of interstellar gas, terminating star formation in a very short timescale.
  Since the best-fit SFH of Fornax in the K11 model decreases rapidly in the first short epoch (i.e., $\lesssim 0.1$~Gyr), $A_\ast$ is considered to be a maximum in their model.
  On the other hand, since the infall rate in our model is adjusted to keep the interstellar gas mass satisfying the required condition of Equation~(\ref{eq:mass}) for a given SFR, $A_\ast$ can be explored to fit the observed MDF.

  The best-fit parameters of $A_\ast$ and $A_\mathrm{out}$ in the Case~2 model are similar to those in the Case~1 model.
  In the Case~3 model, in which $\alpha$ is changed from the best-fit values in K11 into unity, the best-fit parameters of $A_\mathrm{out}$ are similar to those of Case~2 but those of $A_\ast$ become lower.
  In other words, $A_\ast$ and $\alpha$ are degenerate.

  The best-fit values of $A_\ast$ change dramatically in the Case~4 model, in which the SFH timescales are modified into those used in our model of $\sim 10$~Gyr; $A_\ast$ decreases more than one order of magnitude while $A_\mathrm{out}$ is similar to that in the Case~3 model except for Sculptor\footnote{The best-fit value of $A_\mathrm{out}$ for Sculptor in the Case~4 model has a large uncertainty because of a large $F_\mathrm{ex}$.}.
  The significant decrease of $A_\ast$ can be interpreted as follows.
  In order to reproduce a given MDF, a longer star formation timescale requires a slower chemical evolution, that is, a lower value of $A_\ast$.
  As a consequence, the Case~4 model provides a set of the best-fit parameters which is quite similar to that of our model as shown in Table~\ref{tab:cases}.
  The tiny differences between the two sets of the best-fit parameters are caused by differences in the shapes of SFH.

  Therefore, we can conclude that the difference in the best-fit parameters of $A_\ast$ between the K11 and our model is originated from the difference in the star formation timescale for dSph.
  Furthermore, although \citet{Fenner_06} made their model fitting for the observed chemical abundance of Sculptor and found that the $\mathrm{[\alpha/Fe]}$ is degenerated between the duration of the SFH and the SN feedback, our model fitting for the observed MDFs show no degeneracy between the duration of the SFHs and the $M_\mathrm{out}$ (see the Cases 3 and 4 in Table 3).
  We emphasize that, since the parameters in chemical evolution model are sensitive not only to the MDF but also to the SFH, it is important to fit the MDF and SFH simultaneously in order to investigate the chemical evolution of dSph.

  \subsection{The Discrepancy of the Minimum Delay Time for SNe~Ia between Our Model Results and Observations}\label{subsec:DiscrepTdelaymin}

  In Section~\ref{subsec:Param+Errors}, we show that our model disfavors the minimum delay time of $t_\mathrm{delay, min} \sim 0.1$~Gyr since it is too short to reproduce the observed distribution of the stars in our sample dSphs in the $[\mathrm{Mg/Fe}]$--$[\mathrm{Fe/H}]$ plane (see right panels of Figure~\ref{fig:MDF}).
  This is against the recent observational and/or theoretical estimates for $t_\mathrm{delay,min}$, which show that $t_\mathrm{delay,min} \lesssim 0.1$~Gyr (e.g., \citealt{Totani_08, Cooper_09, Maoz_12, Kistler_13}).
  While the K11 model with $t_\mathrm{delay, min} = 0.1$~Gyr successfully reproduces the observed distributions in the $[\mathrm{Mg/Fe}]$--$[\mathrm{Fe/H}]$ plane, it results in too short star formation timescales compared to those estimated observationally for the dSphs.
  Therefore, to explain both of the SFH and the chemical abundance of dwarf galaxies simultaneously, a longer $t_\mathrm{delay,min}$ than the observationally favorable value of $\sim 0.1$~Gyr seems to be required inevitably.
  What are the origins of this discrepancy in $t_\mathrm{delay,min}$ between our chemical evolution model and the observational and theoretical estimates?

  One of the origins of this discrepancy can be the dependence of SN~Ia rate on metallicity.
  Taking account of the suggestion from some theoretical works that the SN~Ia rate in metal-poor galaxies is smaller compared to metal-rich galaxies \citep{Kobayashi_98, Kobayashi_09, Meng_11}, we consider that the effect of the SN~Ia yields on the abundance ratio of $[\mathrm{Mg/Fe}]$ will be smaller in the early metal-poor stages of chemical evolution than in the late metal-rich stages.
  Incorporating such metal-dependent SN~Ia rate into our model with short $t_\mathrm{delay,min}$ will result in the same abundance ratios as the model with long $t_\mathrm{delay,min}$ because additional time is required to emerge the effects of SN~Ia on abundance ratios since the onset time of the first SN~Ia.
  Therefore, the discrepancy can be relaxed if such metallicity dependence of SN~Ia rate is taken into account.
  However, the opposite dependence on metallicity has also been suggested (e.g., \citealt{Cooper_09, Kistler_13}).
  Therefore, it is still unclear how $t_\mathrm{delay, min}$ depends on the metallicity of the system.

  Another possible origin of this discrepancy can be the one-zone approximation adopted in our model; it is an apparently oversimplified approximation for the dSphs because the age and metallicity gradients have been observed in them as described in Section~\ref{sec:4-2}.
  If we adopt a multi-zone approximation instead of the one-zone one, the discrepancy can be relaxed.
  While the one-zone model sets the system chemically homogeneous, multi-zone models probably can describe these complicated structures.
  In the multi-zone approximation, zones can have different SFHs with each other and the ejecta of SN~II and SN~Ia can flow from a zone to another.
  If a zone has a  short ($< 0.1$~Gyr) star formation event and its star formation ceases before the SN~Ia ejecta from near zones flows in the zone, the chemical abundance of stars formed there reflect only the SN~II ejecta.
  In other words, the chemical evolution model with multi-zone approximation can increase the metallicity of the stars without the effect of SN~Ia.

  The assumption of the time-independent IMF is also one of the origins of the discrepancy, since some theoretical works predicted a different IMF when the metallicity of the system is low.
  \citet{Tumlinson_06} suggested that low-mass star formation was inhibited when the metallicity of the system was below a critical metallicity (i.e., $Z_{\rm cr} \lesssim 10^{-4}~Z_\odot$).
  If the system forms only massive stars when the metallicity of the ISM is low, the ISM is chemically enriched without the contribution of SNe~Ia.
  Therefore, the metallicity-dependent IMF can relax the discrepancy about $t_{\rm delay,min}$.
  
  Some previous studies argue that dSphs have a bottom-heavy IMF and/or lack very massive ($\ge 25 M_\odot$) stars \citep{Tsujimoto_11, Li_13, Weidner_13}.
  \citet{Li_13} analysed the chemical abundance of the stars in Fornax dSph and concluded that a bottom-heavy IMF is necessary to explain the low $\mathrm{[\alpha/Fe]}$.
  However, the low $\mathrm{[\alpha/Fe]}$ is also explained by the large contribution of the SN~Ia (\citealt{Ikuta_02, Lanfranchi_04, Tolstoy_09}; K11), and a bottom-heavy IMF cannot explain the high $\mathrm{[\alpha/Fe]}$ (e.g., $\mathrm{[Mg/Fe]} > 0.5$), such as observed in our samples.
To confirm the effect of the bottom-heavy IMF on our model results, we calculate the chemical evolution of the samples by changing the stellar mass upper limit from 100~$M_\odot$ to 25~$M_\odot$.
  However, we have found that the model MDFs have no changes and the model $\mathrm{[Mg/Fe]}$ decrease with decreasing mass upper limit at the same $\mathrm{[Fe/H]}$ about 0.2~dex.
  Therefore, the model with a bottom-heavy IMF and $t_\mathrm{delay,min} = 0.1$~Gyr results in worse fitting to the observed data.

  \subsection{The Chemical Evolution Histories of dSphs Indicated by Our Best-fit Models}\label{sec:4-4}

   \subsubsection{The Star Formation Efficiencies of Dwarf Galaxies}\label{subsubsec:SFE}

   As shown in Table~\ref{tab:results}, the best-fit values of the SFE ($A_\ast$) for our sample dSphs are $10^{-2.4}$--$10^{-1.0}~\mathrm{Gyr^{-1}}$.
   Here we compare these model results with the observational estimates for the SFEs of dwarf galaxies.
   Note that the SFE derived in our model is constant during the entire history of the galaxy while the SFE estimated from a star-forming dwarf galaxy reveals its current value.

   \begin{figure}
    \begin{center}
     \plotone{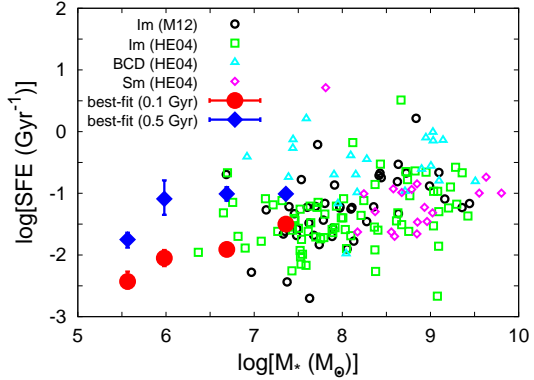}
     \caption{Comparison of the model results with observed data of star-forming dwarf galaxies in the SFE--stellar mass plane.
	 The model results are plotted as filled symbols while the observed data are plotted as open symbols.
	 The filled circles and diamonds with error bars are the results of the best-fit models with $t_\mathrm{delay,min} = 0.1$~Gyr and 0.5~Gyr, respectively.
	 The open circles are the data of Im galaxies estimated from gas and stellar masses in \citet{McCall_12} and H$\alpha$ luminosities in \citet{Kennicutt_08}.
	 The open squares, triangles, and diamonds are the data of Im, BCD, and Sm galaxies, respectively, estimated from gas masses and H$\alpha$ luminosities in \citet{Hunter_04} and stellar masses derived from \citet{Hunter_06} and \citet{Bell_01}.\label{fig:datas_As}}
    \end{center}
   \end{figure}
   Figure~\ref{fig:datas_As} shows such a comparison of the SFE as a function of the stellar mass between our model results and the observational estimates.
   In this plot, the observational estimates for the SFEs of the star-forming dwarf galaxies are shown instead of dSphs since the star-forming dwarf galaxies have enough amount of interstellar gas and SFR to be detected by observations, which are inevitable to evaluate current SFE.
   From 48 Im galaxies whose \ion{H}{1} gas masses and H$\alpha$ luminosities are obtained by \citet{McCall_12} and \citet{Kennicutt_08}, respectively, the mean SFE is evaluated as $\langle \log{[\mathrm{SFE} / \mathrm{Gyr^{-1}}]} \rangle = -1.4\pm 0.7$, which is consistent with our best-fit SFEs.
   The mean SFEs of Im and Sm galaxies in \citet{Hunter_04} are $\langle \log{[ \mathrm{SFE} / \mathrm{Gyr^{-1}}]} \rangle = -1.4\pm 0.5$ and $-1.1\pm 0.5$, respectively, which are also comparable to those of our best-fit models.
   The mean SFE of blue compact dwarf (BCD), $\langle \log{[ \mathrm{SFE} / \mathrm{Gyr^{-1}}]} \rangle = -0.6\pm 0.5$, is found to be higher by almost one magnitude than our best-fit SFEs.

   The result of this comparison of the SFE is consistent with the observed result that the mean SFHs of the different morphological types of dwarf galaxies including dSph and dIrr are indistinguishable with each other over most of cosmic time \citep{Weisz_11}.
   Therefore, the dIrrs may experience the same chemical evolution histories as the dSphs essentially.
   It is indicated that, if dIrrs finish up their all interstellar gas, they can evolve into dSphs \citep{Kormendy_85, Peeples_08, Zahid_12}.

   \subsubsection{The Gas Outflow from dSphs}

   As described in Section~\ref{sec:3-2}, our best-fit parameter values of $A_\mathrm{out}$ for our sample dSphs are found to be significantly large; the total masses of outflow gas are 10--100 times larger than the final stellar masses of the dSphs.
   The best-fit parameter values of $A_\mathrm{out}$ themselves and/or the result of a significant amount of outflow gas are consistent with the results from other analytic models for chemical evolution (\citealt{Lanfranchi_03, Lanfranchi_04, Lanfranchi_07, Lanfranchi_10, Lanfranchi_06a, Lanfranchi_06b, Lanfranchi_08}; K11). 
   This amount of outflow gas mass is consistent with the physically possible maximum value, $M_\mathrm{out}^\mathrm{max}$, estimated from the following simple consideration; that is, the 100\% of the SN energy is converted into the kinetic energy of outflow gas with a typical escape velocity of $\sim 25~\mathrm{km~s}^{-1}$ (Sculptor; \citealt{Sanchez-Salcedo_07}), resulting in $\dot{M}_\mathrm{out}^\mathrm{max} \sim 100 \Psi$ or $M_\mathrm{out}^\mathrm{max} \sim 100 M_\star$.
   Nevertheless, this amount of outflow gas mass significantly depends on the assumption regarding the metallicity of outflow gas.
   The low metallicity of the dSphs is achieved not by the amount of gas outflow but by the amount of ejected metals.

   In our best-fit models with $t_\mathrm{delay,min} = 0.1$~Gyr (0.5~Gyr), the mass fractions of Fe ejected from the system to the totally synthesized Fe in the system are evaluated as 95\% (91\%), 99\% (98\%), 99\% (98\%), and 99\% (99\%) for Fornax, Sculptor, Leo~II, and Sextans, respectively; that is, more than 90\% of Fe is lost from all of our sample dSphs as outflow powered by superwind, whose amount is consistent with those of \citet{Kirby_11b}.
   This high ejected mass fraction of Fe is inevitably required to reproduce the observed low metallicities of the dSphs.
   In our model, 100\% of the SN ejecta is assumed to be mixed instantaneously within the ISM at once before gas outflow occurs; the SN ejecta are diluted with the entire interstellar gas and the metallicity of outflow gas becomes significantly smaller than that of the SN yields.

   High resolution chemo-dynamical simulations suggest that only a low fraction of SN ejecta stays in the region where star forms \citep{Marcolini_06, Revaz_12}.
   For example, according to the results from \citet{Revaz_12} that the SN feedback efficiency of 0.03--0.05 is sufficient to reproduce the properties of the dSphs, the mass fraction of the SN ejecta mixing with the ISM is estimated to be only $\sim 5$\%.
   In this estimation, the total mass of outflow gas results in only $\sim 0.1$\% of the final stellar masses for all of our sample dSphs.
   Therefore, the total mass of outflow gas can vary about 5 orders of magnitude according to the assumption of the metallicity of outflow gas.
   This result indicates that the ejected metal mass fraction of $> 90$\% is more essential in the chemical evolution history of dSphs than the total mass of outflow gas.

   In order to constrain the total mass of outflow gas, it is important to evaluate what fraction of SN ejecta is directly expelled from the system.
   However, it is strikingly difficult to observe the outflow from dSphs since they have almost no gas at present.
   In contrast, the outflowing gas has been observed in dIrrs \citep{Heckman_01, Martin_02, van_Eymeren_09a, van_Eymeren_09b, van_Eymeren_10}.
   For the dIrr starburst galaxy NGC~1569, \citet{Martin_02} estimated the metal content of outflow gas by using {\it Chandra} and concluded that the outflow transports most of the metals synthesized by current starburst.
   If the dIrrs are considered to be the progenitors of the dSphs as described in Section~\ref{subsubsec:SFE}, which still have plentiful gas and show ongoing star formation activity, these observational results imply that a significant fraction of SN ejecta is also directly expelled from dSphs.
   Therefore, the total masses of outflow gas of dSphs seem to be smaller than their current stellar masses.

 \section{Summary}\label{sec:summary}

 We have constructed a new analytic model for the chemical evolution to reproduce both the observed SFH and MDF of a dSph, simultaneously.
 While the prescriptions and assumptions in our model are similar to those adopted in K11, there are critical differences between them as described in Section~\ref{sec:assumption}: that is, the assumptions for the SFH, the minimum delay time for SN~Ia, and infalling gas.
 With this new model, we have performed model calculations for Fornax, Sculptor, Leo~II, and Sextans, and compare the derived MDFs with observed MDFs.
 Our results and conclusions are summarized as follows.

 \begin{enumerate}
  \item It is demonstrated that our new models can reproduce the observed MDFs of the dSphs with various metallicity.
  We have found that their MDFs are characterized by two parameters: the SFE ($A_\ast$) and the gas outflow efficiency ($A_\mathrm{out}$).
  \item It is shown that the likelihoods of the best-fit models for the dSphs do not depend on the minimum delay time of SN~Ia ($t_\mathrm{delay,min}$).
  This means that any constraints are not obtained by analyzing both the MDFs and the SFHs.
  We have found that the best-fit models with $t_\mathrm{delay,min} = 0.1$~Gyr underestimate the observed $\mathrm{[Mg/Fe]}$ of the dSphs, while previous studies have suggested $t_\mathrm{delay,min} \simeq 0.1~\mathrm{Gyr}$.
	In our models, $t_\mathrm{delay,min}$ is required to be much longer than $0.1$~Gyr to explain the observed SFHs, MDFs, and $\mathrm{[Mg/Fe]}$, simultaneously.
  \item The derived SFEs of the dSphs are lower than that estimated observationally for the MW ($\sim 0.24~\mathrm{Gyr}^{-1}$) and those obtained theoretically for the dSphs by K11.
  These low SFEs lead to small growth rate of metallicity and broad metal-poor tail in the MDF.
  The best-fit values of SFE are similar to those evaluated observationally for dIrrs with both rich gas and star formation activity.
  This is consistent with the observations of dwarf galaxies that the mean SFHs of the dSphs and dIrrs are indistinguishable with each other over most of cosmic time.
  \item The derived $A_\mathrm{out}$ (the outflow gas mass per unit SN) are similar to those obtained theoretically for the dSphs by K11 and the outflow rate per SFR are larger than that estimated observationally for the MW ($\sim 0.5$).
  The large amount of gas outflow leads to large amount of metal outflow and lower mean metallicity of the dSph.
  Our model predicts that the dSphs have lost more than 90\% of iron synthesized in the systems; this is consistent with observational results that there is outflowing gas in some dIrrs like NGC~1569.
  \item The derived gas infall rates are comparable to the derived gas outflow rates.
  Since the dSphs have low metallicities and relatively long SFHs, our model needs large amount of metal outflows and as well as comparable amount of gas infalls to maintain the star formations as observed.
 \end{enumerate}

\acknowledgments

We are grateful to M.~Chiba, K.~Hayashi, D.~Toyouchi, T.~Otsuka, and T.~Obata for insightful discussions and comments.
We would like to thank the referee for his/her useful comments and suggestions.
This work was financially supported in part by the Japan Society for the Promotion of Science (No.23244031 [YT]).

\appendix

\section{THE EFFECT OF THE DIFFERENT SIZE OF TIME STEP ON OUR MODEL RESULTS}

 \begin{figure}
  \begin{center}
   \plottwo{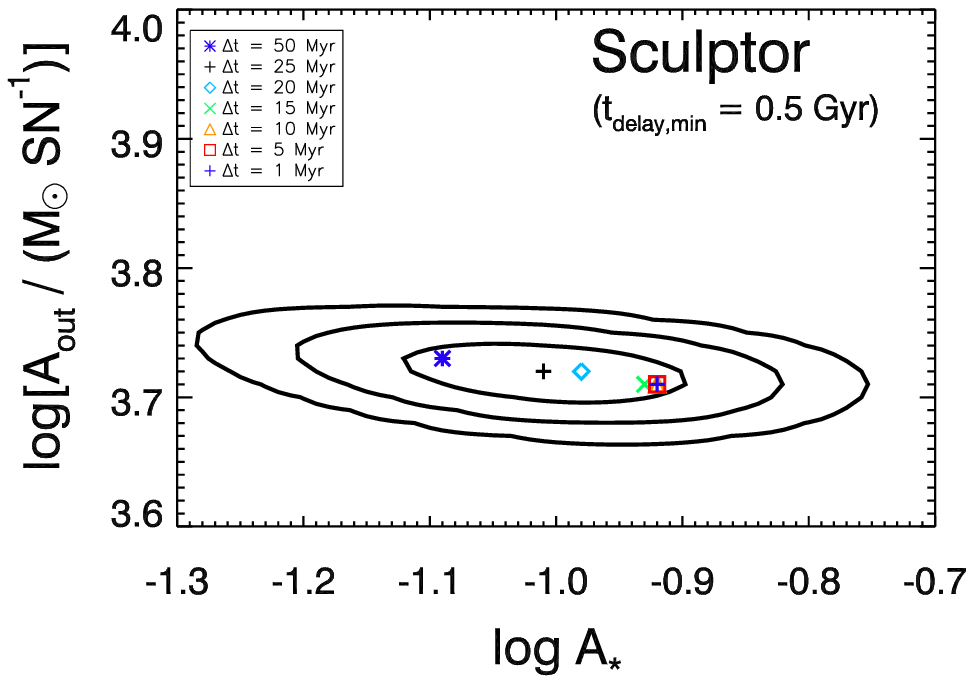}{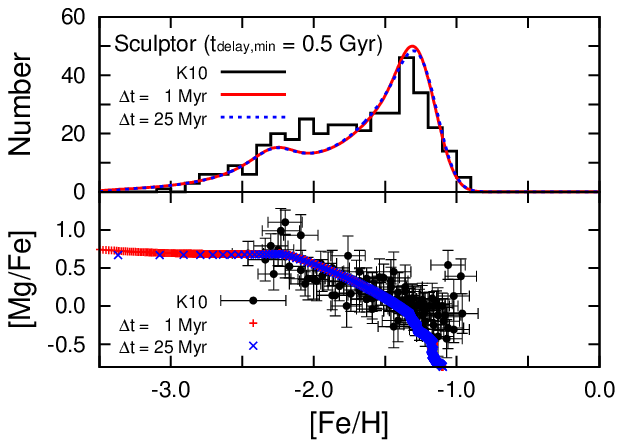}\\
   \caption{(Left) The best-fit parameters with various sizes of time step and the confidence levels for  the model of Sculptor with $t_\mathrm{delay,min}=0.5$~Gyr.
   The black plus symbol and the contours are exactly the same as those in Figure~\ref{fig:contour}, that is, the best-fit parameter with $\Delta t = 25$~Myr and the $1\sigma$, $2\sigma$, and $3\sigma$ confidence levels, respectively.
   The best-fit parameters with $\Delta t = 50$, 20, 15, 10, 5, and 1 Myr are plotted as blue asterisk, cyan diamond, green cross, orange triangle, red square, and blue plus, respectively.
   (Right) The MDFs and the $\mathrm{[Mg/Fe]}$--$\mathrm{[Fe/H]}$ relations of Sculptor.
   The histogram and the data points with error bars are exactly the same as those in Figure~\ref{fig:MDF}, that is, the observed MDF and observed chemical abundance of stars in Sculptor, respectively.
   Red solid curve and Blue dotted curve are the model MDFs, which are calculated by using the best-fit parameter set of Sculptor with $t_\mathrm{delay,min}=0.5$~Gyr as listed in Table~\ref{tab:results}, with different size of time step of $\Delta t = 1$~Myr and 25~Myr, respectively.
   Red pluses and blue crosses are also the model calculated by the same parameter set with $\Delta t = 1$~Myr and 25~Myr, respectively, they are plotted at each time step.\label{fig:time_step}}
  \end{center}
 \end{figure}

 We adopted a time step of 25~Myr in our calculations (Section \ref{sec:assumption}).
 This step is much longer than that of K11 ($\Delta t_\mathrm{K11}=1$~Myr).
 In this section we examine whether the time step affects the results or not.

 For the four dSphs, we evaluate the best-fit parameters for time steps of $\Delta t = 50$, 20, 15, 10, 5, and 1~Myr for the parameter sets which sufficiently cover the areas of $1\sigma$ confidence level shown in Figure~\ref{fig:contour}.
 As a result, it is confirmed that the $1\sigma$ confidence regions for $\Delta t=25$~Myr include all the best-fit parameter sets for $\Delta t = 1$--50~Myr for all the sample dSphs with $t_\mathrm{delay, min} = 0.5$~Gyr and 0.1~Gyr.
 In Figure~\ref{fig:time_step} we show the result for Sculptor with $t_\mathrm{delay, min} = 0.5$~Gyr as an example.
 The best-fit parameter set only slightly changes by changing the time step as shown in the left panel of Figure~\ref{fig:time_step}.

 The difference in the time step affects the chemical abundance of the ISM mainly at an early phase.
 In the right panel of Figure~\ref{fig:time_step}, we show the calculated MDF and the $\mathrm{[Mg/Fe]}$--$\mathrm{[Fe/H]}$ relation for the best-fit models of Sculptor ($t_\mathrm{delay, min} = 0.5$~Gyr) with $\Delta t = 25$, and 1~Myr.
 Note that the calculated $\mathrm{[Mg/Fe]}$--$\mathrm{[Fe/H]}$ relation of each time step is shown by symbols.
 There is no significant difference between the model MDFs for $\Delta t = 1$~Myr and $\Delta t = 25$~Myr. 
 At the early phase (i.e., $\mathrm{[Fe/H]} < -2.5$), the values of $\mathrm{[Mg/Fe]}$ for $\Delta t = 1$~Myr are slightly larger than those for $\Delta t = 25$~Myr.
 Since the more massive stars have the shorter lifetimes, the chemical abundance of the ISM at the early phase depends on the yields of massive stars.
 In the first time step of calculations, stars with mass of $m \geq 10~M_\odot$ die within 25~Myr while those with $m \geq 100~M_\odot$ dies within 1~Myr.
 In other words, the model with $\Delta t = 25$~Myr cannot resolve the yields of stars with $m \geq 10~M_\odot$ while the model with $\Delta t = 1$~Myr can treat their yields each other.
 However, the difference of this effect is small and negligible as metallicity increases.
 Therefore, the effect of the different time step on our model results is negligible.

 \section{THE ABUNDANCE RATIOS OF $\mathrm{[Si/Fe]}$, $\mathrm{[Ca/Fe]}$, AND $\mathrm{[Ti/Fe]}$}

 \begin{figure}
  \begin{center}
   \plotone{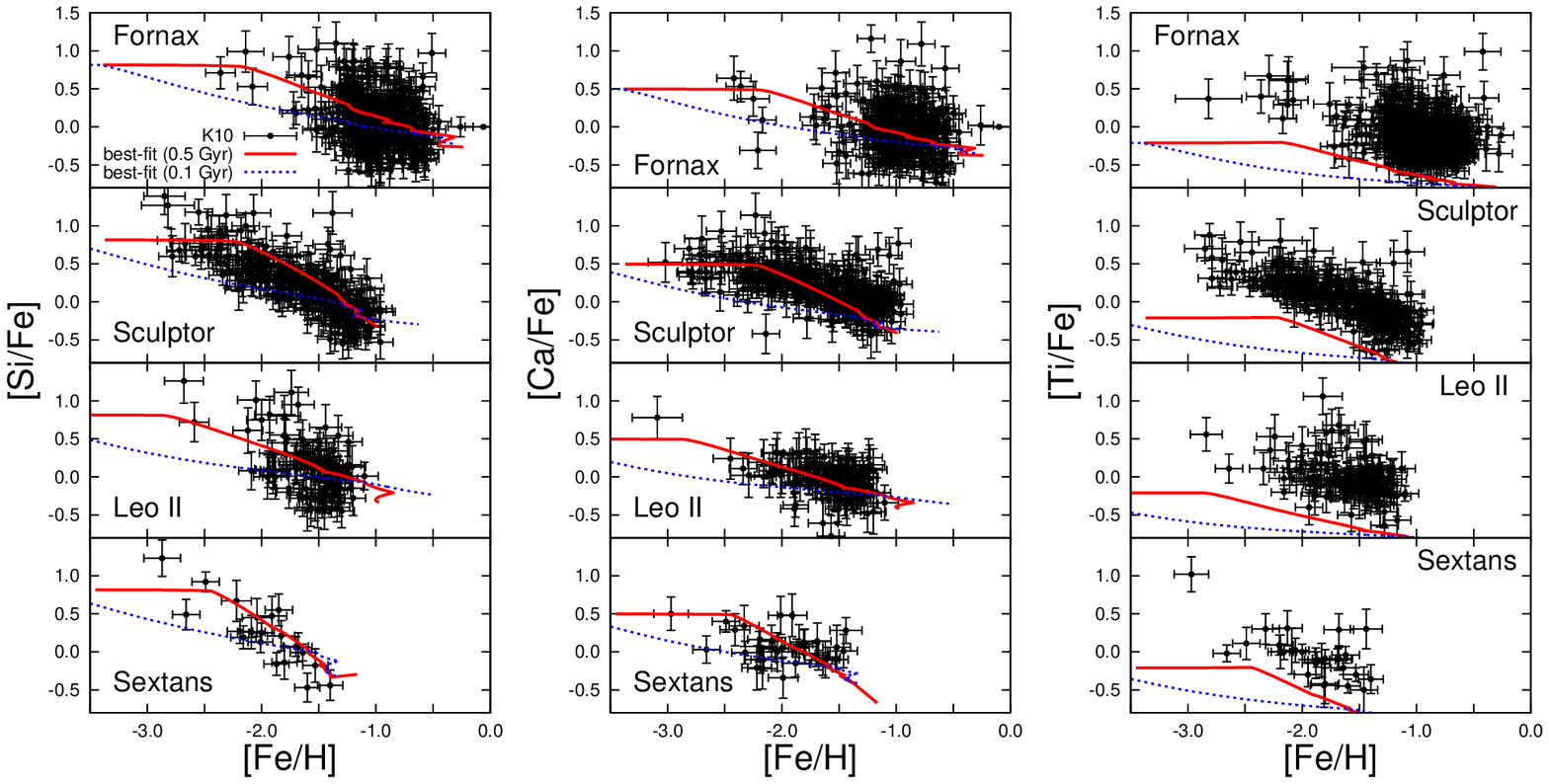}
   \caption{The chemical abundance ratios of $\mathrm{[Si/Fe]}$ (left panels), $\mathrm{[Ca/Fe]}$ (middle panels), and $\mathrm{[Ti/Fe]}$ (right panels).
   The data points with error bars are taken from \citet{Kirby_10}.
   Same as the right panel of Figure~\ref{fig:MDF}, only observed data with uncertainties less than or equal to 0.3~dex are shown.
   The blue dotted and red solid curves are the results of our best-fit models with $t_\mathrm{delay,min} = 0.1$~Gyr and 0.5~Gyr, respectively.\label{fig:abundance}}
  \end{center}
 \end{figure}

 Here we present the evolution of Si, Ca, and Ti, which are also shown in K11.
 As shown in Figure \ref{fig:abundance}, the best-fit models with $t_\mathrm{delay,min} = 0.5$~Gyr result in better fits to the observed data of $\mathrm{[Si/Fe]}$ and $\mathrm{[Ca/Fe]}$ provided in \citet{Kirby_10} than those with $t_\mathrm{delay,min} = 0.1$~Gyr.
 This result is consistent with the result described in Section~\ref{subsec:Param+Errors} and shown in the right panel of Figure~\ref{fig:MDF}; that is, $t_\mathrm{delay,min} = 0.1$~Gyr is too short to reproduce the observed data of the dSphs.

 On the other hand, the best-fit models with both $t_\mathrm{delay,min} = 0.1$~Gyr and 0.5~Gyr underpredict the observed data of [Ti/Fe] as shown in the right panel of Figure \ref{fig:abundance}.
 Such underprediction is also seen in K11 and they ignored the model result for Ti since they consider that the yield for $\mathrm{[Ti/Fe]}$ is inaccurate.
 Therefore, this underprediction is simply because the SN~II yield for $\mathrm{[Ti/Fe]}$ used in our model \citep{Nomoto_06} is not high enough to reproduce the observed data.

\end{document}